\documentclass[aip,jcp,preprint,amsmath,amssymb, groupedaddress]{revtex4-1}
\usepackage[T1]{fontenc}
\usepackage[utf8]{inputenc}
\usepackage{amsfonts}
\usepackage{multirow,dcolumn}
\usepackage{booktabs}
\usepackage[version=3]{mhchem}
\usepackage{setspace} 
\usepackage{lmodern}
\usepackage{leftidx}

\usepackage{caption}
\usepackage{tabularx}
\usepackage{threeparttable}
\usepackage[dvipsnames]{xcolor}
\usepackage{blindtext}
\usepackage{booktabs}
\usepackage[version=3]{mhchem}
\usepackage{lmodern}
\usepackage{natbib}
\usepackage[pdftex]{graphicx}
\usepackage[amssymb]{SIunits} 

\usepackage{paralist} 
\usepackage[normalem]{ulem}

\newcommand{\TT}{\mathcal{T}}
\newcommand{\SSS}{\mathcal{S}}
\newcommand{\dg}[1]{#1^\dagger}
\newcommand{\dgg}{^\dagger}

\newcommand{\etd}{e^{T^\dagger}}
\newcommand{\etdm}{e^{-T^\dagger}}
\newcommand{\esd}{e^{S^\dagger}}
\newcommand{\esdm}{e^{-S^\dagger}}
\newcommand{\esm}{e^{-S}}
\newcommand{\etm}{e^{-T}}
\newcommand{\es}{e^{S}}
\newcommand{\et}{e^{T}}

\newcommand{\xmsr}{X - \srx}

\newcommand{\rd}[1]{{\color{black}#1}}

\newcommand{\xmz}{X_0}

\newcommand{\cm}[2]{ [ #1,#2]}

\newcommand{\sump}{\sideset{}{'}\sum}

\newcommand{\pz}{\Psi_0}
\newcommand{\phm}{\hat{\mathcal{P}}}

\newcommand{\mian}{\langle e^T|e^T\rangle}

\newcommand{\omzy}{\Omega_0^Y}
\newcommand{\omzz}{\Omega_0^Z}

\newcommand{\omy}{\Omega^Y(\omega_Y)}
\newcommand{\omz}{\Omega^Z(-\omega_Z)}

\newcommand{\brakm}[2]{\frac{\left \langle #1 |#2 \right\rangle}{\mian}}
\newcommand{\sred}[1]{\langle #1\rangle}

\newcommand{\fr}[1]{Eq.~(\ref{#1})}

\newcommand{\Frf}[1]{Fig.~(\ref{#1})}
\newcommand{\Frt}[1]{Table~\ref{#1}}

\newcommand{\Frttt}[2]{Tables~\ref{#1}~and~\ref{#2}}
\newcommand{\frs}[1]{section~\ref{#1}}

\newcommand{\srx}{\langle X \rangle}

\newcommand{\quadra}{\langle\langle X; Y, Z \rangle\rangle}
\newcommand{\quadraom}{\langle\langle X; Y, Z \rangle\rangle_{\omega_Y, \omega_Z}}

\newcommand{\brakett}[3]{\left\langle #1\middle| \right. #2 \left. \middle|#3\right\rangle} 

\newcommand{\quadru}[3]{\frac{\brakett{\Psi_0}{#1}{K} \brakett{K}{#2-\brakett{\Psi_0}{#2}{\Psi_0}}{N} \brakett{N}{#3}{\Psi_0}}{(\omega_K +\omega_{#1})(\omega_N -\omega_{#3})}}

\newcommand{\bb}[2]{\big\langle #1\big|#2\big\rangle} 

\newcommand{\braket}[2]{\left\langle #1\middle|#2\right\rangle}

\newcommand{\equ}[1]{\begin{equation} #1 \end{equation}}
\newcommand{\equl}[2]{\begin{equation}\label{#2} #1 \end{equation}}

\newcommand{\equal}[2]{\begin{align}\label{#2} #1 \end{align}}
\newcommand{\equs}[1]{\begin{equation}\begin{split} #1 \end{split}\end{equation}}
\newcommand{\equsl}[2]{\begin{equation}\begin{split}\label{#2} #1 \end{split}\end{equation}}

\begin{document}

\title{Transition moments between excited electronic states from the Hermitian formulation of the coupled cluster quadratic response function}

\author{Aleksandra M. Tucholska}
\email{tuchol@tiger.chem.uw.edu.pl}
\affiliation{Faculty of Chemistry, University of Warsaw, Pasteura 
  1, 02-093 Warsaw, Poland}
\author{Michał Lesiuk}
\affiliation{Faculty of Chemistry, University of Warsaw, Pasteura 
  1, 02-093 Warsaw, Poland}
\author{Robert Moszynski}
\affiliation{Faculty of Chemistry, University of Warsaw, Pasteura 
  1, 02-093 Warsaw, Poland}
\affiliation{Kavli Institute for Theoretical Physics, University of California, Santa Barbara, CA 93106-4030, USA}

\begin{abstract}

We introduce  a new method for the  computation of the transition moments between the excited electronic states based
on the expectation value formalism of  the coupled cluster theory \rd{(XCC)} [B. Jeziorski and R. Moszynski, 
Int. J. Quant. Chem. {\bf 48}, 161 (1993)]. The working expressions of the new method solely
employ the coupled cluster \rd{operator $T$ and an auxiliary operator $S$ that is expressed as a finite commutator
expansion in terms of $T$ and $T^\dagger$}.
In the approximation adopted in the present paper the cluster expansion is limited to single, double, and linear triple excitations.
The computed dipole transition probabilities for the  singlet-singlet and triplet-triplet
transitions in alkali earth atoms agree well with the available theoretical and experimental data.
In contrast to the existing coupled cluster response theory, the matrix elements
obtained by using our approach satisfy the Hermitian symmetry even if the excitations in the cluster operator are truncated, \rd{but the operator $S$ is exact.
The Hermitian symmetry is slightly broken if the commutator series for the operator $S$ are truncated.} 
As a part of the numerical evidence for the new method, we report calculations of the transition moments between the excited triplet states which have not yet been reported in the literature within the coupled cluster theory.
Slater-type basis sets constructed according to the correlation-consistency principle are used in our calculations.

\end{abstract}
\begin{center}
\maketitle
\end{center}

\section{Introduction}

Response of a system to external perturbations is described by linear,
quadratic, and higher-order response functions.\cite{zubarev1974nonlinear, linderberg2004propagators,
oddershede1987propagator} 
Many physical observables such as transition probabilities,
dynamic polarizabilities, hyperpolarizabilities, and lifetimes are defined through the response functions or 
can be derived from the response functions.
Until recently, properties of the excited electronic states were not
easily available in high-resolution experiments, but with the advances of
new spectroscopic techniques in the hot pipe
\cite{rybak2011femtosecond, rybak2011generating, amaran2013femtosecond, levin2015coherent, wilson2015femtosecond}
and ultracold experiments,\cite{skomorowski2012rovibrational, mcguyer2015precise,
  mcdonald2016photodissociation, mcguyer2013nonadiabatic, mcguyer2015control} more and
more accurate experimental data become available and possibly need
theoretical interpretation. Theoretical information about the transition
moments between the excited states is also necessary to propose new routes
to obtain molecules in the ground rovibrational state (see, e.g., Ref. \onlinecite{tomza2012optimized}).
Last but not least, excited states properties define the asymptotics of the
excited state interaction potentials,\cite{skomorowski2011long} and play an
unexpectedly important role in the dynamics of nuclear motions in the
presence of external fields. \cite{tomza2013interatomic}

The properties of the excited states, e.g., polarizabilities, transition strengths, and lifetimes can be obtained from limited multiconfiguration interaction theory, but this approach inherently suffers from the
size inconsistency problem. Applying the size consistent coupled cluster (CC) formalism to the response function  opens up a possibility of an accurate description of molecular properties with an affordable computational cost for medium size molecules. In the 1990s J\o rgensen and
collaborators formulated the CC response theory,\cite{koch1990coupled, koch1997coupled} based on the coupled cluster generalization of the Hellmann-Feynman theorem where the average value is replaced by a transition expectation value with respect to the coupled cluster state. 
However, in this theory the necessary Hermiticity condition required from the
transition moments is not satisfied, and in some cases this  leads to unphysical numerical results.

In the present study we focus on the molecular properties
that can be obtained from the quadratic response function,
$\quadraom$. The latter describes the response of an observable $X$ to perturbations $Y$ and $Z$
oscillating with the frequencies $\omega_Y$ and $\omega_Z$, respectively. \rd{In the exact case} the transition moment $\TT_{LM}^X$  
between the excited states \rd{$L$ and $M$ can be} computed from the  double residue of the
quadratic response function
\equl{\lim_{\omega_Y\rightarrow -\omega_L}(\omega_L + \omega_Y)\lim_{\omega_Z\rightarrow \omega_M}(\omega_M - \omega_Z) \quadraom  = \TT^Y_{0L} (\TT^X_{LM}-\delta_{LM}\brakett{\Psi_0}{X}{\Psi_0})\TT^Z_{M0},}{limlim}
where $\TT^Y_{0L}$ and $\TT^Z_{M0}$ are transition
moments between the ground and excited sates, and $\omega_K$
is the excitation energy of the state $K$.  
\rd{Note that the Kronecker delta term $\delta_{LM}$ appearing in the above expression is responsible for the
cancellation of the disconnected terms in the quadratic response function as in the standard third-order
perturbation theory. When $L\ne M$, and this is always the case, this term simply vanishes.}
For different $L$ and $M$ states the
  transition strength  $\SSS_{LM} $ is defined as
   \equl{\SSS_{LM} = |\TT_{LM}|^2.}{ts}
The transition moments are necessary to compute
the transition probabilities\cite{drake2006springer}
\equl{A_{LM} = \frac{1}{3} \frac{16\pi^3}{3h\epsilon_0\lambda^3}\SSS_{LM},}{tpr}
where $\epsilon_0$ is the vacuum permittivity, $\lambda$ is the wavelength, $h$ is the Planck constant,
and  $S_{LM}$ is the transition strength.
The  lifetime\cite{drake2006springer} of a state $L$ is defined as
\equl{\tau_L = \frac{1}{\sum_KA_{LK}}.
}{lif}

There exists two coupled cluster approaches for the computation of the transition moments between the ground and excited states, the
linear response coupled cluster theory (LRCC) of Koch et al.\cite{koch1990coupled, koch1994caclculation, koch1997coupled, christiansen1998response}
and the coupled cluster expectation value formulation of the linear response function (XCC) of Tucholska et al.\cite{tucholska2014transition}
As already stated above, for the transition moments between the excited states, the only available approach is based on the quadratic response coupled cluster
(QRCC) theory of Koch et al.\cite{koch1990coupled, koch1994caclculation, koch1997coupled, christiansen1998response}
In the present work we generalize the approach of Refs.~\citenum{moszynski2005time} and \citenum{tucholska2014transition} to the calculation
of transition properties between the excited states.
The transition moments, $\TT_{LM}^X$, where $L$ and $M$ denote the singlet or triplet excited states, are extracted from
the response function to compute lifetimes and transition probabilities.

In the exact theory, the transition moments are Hermitian
\equl{\TT_{LM}^X = (\TT_{ML}^X)^{\star},
}{hermit0}
but this relation is violated 
by the existing QRCC method, in some cases to a large degree, when the cluster operator is truncated at some excitation level.
 In extreme cases this leads to non-physical, negative transition strengths 
 which will be discussed in detail in the remaining part of this work. 
Recently, a new approach to the problem has been proposed,
where molecular properties are computed as derivatives of the eigenvalues of a Hermitian
eigenproblem.\cite{pawlowski2015molecular}
This approach should apparently remove the inaccuracies and inconsistencies of the QRCC theory.
However, numerical results for this method are not yet available and we cannot assess its accuracy. Therefore, we will
restrict our comparisons to the original QRCC theory.

This paper is organized as follows. In sections \ref{basic} and \ref{xcc} we derive the formula for the
XCC transition moments between the excited states. In section \ref{approx} we present the truncations and approximations used
in this work. In  section \ref{num}
we report  numerical results for the transition moments and lifetimes of the Mg and Sr atoms, and for the \ce{Mg2} molecule.
First, we present the comparison of our results with the QRCC method (subsection \ref{comparison2}),
next we compare our results with the available theoretical
and experimental data (subsection \ref{other}) and finally, we investigate the Hermiticity violation in the XCC and QRCC methods (subsection \ref{asd}). In section \ref{concl} we conclude our paper.
\section{Theory}
\subsection{Basic definitions}\label{basic}
In the CC theory the ground state wave function $\Psi_0$ is represented by the exponential Ansatz $\Psi_0 = e^T\Phi$, where the cluster operator $T$ is
given by the 
 sum of $n$-tuple excitation operators $T_n$,
\equ{T = \sum_{n=1}^NT_n,}
\equ{
{T}_n = \frac{1}{n!}\sum_{\mu_n}^N t_{\mu_n} {\mu_n},
}
$\mu_n= E_{ai}E_{bj}\cdots E_{fm}$ is the product of spin-free excitation operators.
$\Phi$ is the Slater determinant built of the occupied orbitals, and $N$ is the number of electrons.
Throughout the work, the indices $a, b, c \ldots$ and $i, j, k \ldots$ denote virtual
 and occupied orbitals,
 respectively,
 and $p, q, r \ldots$ are used in summations over all orbitals. In practical applications the operator $T$ is truncated to make the CC calculation
 computationally feasible. 

The expectation value of an observable $X$ in the XCC theory is given 
by the explicitly connected, size-consistent  expression introduced
 by \citet{jeziorski1993explicitly}
\equl{\frac{\brakett{\et\Phi}{X}{\et\Phi}}{\braket{\et\Phi}{\et\Phi}}=\langle \Phi | e^{S^\dagger}e^{-T}Xe^{T}e^{-S^\dagger}\Phi\rangle.
}{In-av}
\rd{See also the seminal work of \v{C}\'i\v{z}ek \cite{vcivzek1966correlation,vcivzek1969cluster} and other formulations of the CC expectation value problem\cite{noga1988expectation, monkhorst1977calculation, arponen1987extended, arponen1983variational, pal1984variational, pal1985study, pal1986analysis, pal1990coupled}.}
The auxiliary operator $S$ is defined as
\equl{|e^S\Phi\rangle = \frac{|e^Te^{T^\dagger} \Phi\rangle}{\braket{e^T\Phi}{e^T\Phi}}, \qquad S = S_1 + S_2 + \ldots + S_N,}{opers}
and $S_n$ is expressed as\cite{jeziorski1993explicitly}
\equsl{
S_n &= T_n  - \frac1n \hat{\mathcal{P}}_n \left ( \sum_{k=1}
\frac{1}{k!}\cm{\widetilde{T}\dgg}{T}_k \right ) \\
&- \frac1n \hat{\mathcal{P}}_n\left (\sum_{k=1}\sum_{m=0}
\frac{1}{k!}\frac{1}{m!} [\cm{\widetilde{S}}{\dg{T}}_k,T]_m\right),
}{ss}
where \equ{
\widetilde{T} = \sum_{n = 1}^{N} nT_n, \qquad\widetilde{S} = \sum_{n = 1}^{N} nS_n,
}
and $[A, B]_k$ is a $k$-tuply nested commutator.
The superoperator $\hat{\mathcal{P}}_n(X)$ 
yields the $n$-tuple
excitation part of $X$
\equ{\hat{\mathcal{P}}_n(X) = \frac{1}{n!}\sum_{\mu_n} \braket{\widetilde{\mu}_n}{X}{\mu_n},}
where for simplicity we introduce the following notation $\braket{A}{B} = \braket{A\Phi}{B\Phi}$.
\rd{The symbol $\widetilde{\mu_n}$ is used to indicate the use of the biorthonormal basis
$\langle \widetilde{\mu_n}|\nu_m\rangle = \delta_{\mu_n\nu_m}$.
For the single and double excitation manifold we use the basis 
proposed by Helgaker, J\o rgensen, and Olsen, \cite{helgaker2013molecular} and for the triply excited manifold we employ the basis proposed by Tucholska et al.\cite{tucholska2014transition}}

The formula for $S$ is a finite expansion, though it contains terms of high order in the fluctuation potential.\cite{jeziorski1993explicitly}
To find the exact $S$ operator one requires an iterative procedure. However, $S$ can efficiently be approximated while
retaining the size-consistency.
In our previous work,\cite{tucholska2014transition} we presented a hierarchy of approximations and assessed their
accuracy. Let $S_n(m)$ denote the $n$-electron part of $S$, where all available contributions up to the order $m$ in the
fluctuation potential are accounted for. 
In the computations based on the CC3 model (single, double, and linear triple excitations), we employ
\equsl{
S_1(3) &= T_1 +  \hat{\mathcal{P}}_1\left ([T_1^{\dagger}, T_2] \right ),  \\
&+ \hat{\mathcal{P}}_1\left ([T_2^{\dagger}, T_3] \right ),   \\
S_2(3) &= T_2 + \frac12\hat{\mathcal{P}}_2\left ([[T_2^{\dagger}, T_2], T_2] \right ),   \\
S_3(2) &= T_3,   \\
}{s-approx}
where the CC3 equations for  $T_1$, $T_2$ and $T_3$ are given by Koch et al.\cite{koch1997cc3}
\rd{It should be noted that  we take $S_3 = T_3$ from the CC3 theory and no additional terms from \fr{ss}, hence we only include
terms of the  second-order in $S_3$.}
In the instances where the underlying model of the wave function is CCSD (coupled cluster limited to singles and doubles excitations), we employ $S = S_1(3) + S_2(3)$
neglecting the terms including $T_3$.

The exact quadratic response function can be written as the sum over states
\equl{\quadraom = P_{XYZ}
\sum_{\substack{K=1\\N=1}}\quadru{Y}{X}{Z},}{quad1}
where $K$ and $N$ run over all possible excitations, and $|\Psi_0\rangle$ is the ground state.
The action of the permutation operator $P_{XYZ}$ yields six distinct contributions to $\quadra_{\omega_Y,\omega_Z}$ with the indices $X$,
$Y$, and $Z$ being interchanged.
\subsection{XCC transition moments}\label{xcc}
The exact transition moment between the excited states $L$ and $M$ \rd{($L \ne M$)} can be identified from the double residue of 
the quadratic response function\cite{christiansen1998response}
\equsl{&\lim_{\omega_Y\rightarrow -\omega_L}(\omega_L + \omega_Y)\lim_{\omega_z\rightarrow \omega_M}(\omega_M - \omega_Z) \quadraom \\
  &= \brakett{\Psi_0}{Y}{L} \brakett{L}{X-\brakett{\Psi_0}{X}{\Psi_0}}{M} \brakett{M}{Z}{\Psi_0} = \TT^Y_{0L}\TT^X_{LM}\TT^Z_{M0}.}{dble-res}
To obtain $\TT_{LM}^X$ \rd{in XCC theory} we express $\quadra_{\omega_Y,\omega_Z}$ by using the XCC formalism and take the limit of \fr{dble-res}.

Let us introduce the coupled cluster parametrization of the quadratic response function. 
The first order wave function $\Psi^{(1)}(\omega)$ 
is expressible through the resolvent $\mathcal{R}_{\omega}$,
\equ{\Psi^{(1)}(\omega_V) = \mathcal{R}_\omega V|\Psi_0\rangle, \qquad V = Y \text{ or } Z}
\equ{\mathcal{R}_\omega = \sum_{N=1} \frac{|N \rangle \langle N|}{\omega_N + \omega}.}
\rd{Using these definitions,} the expression for the quadratic response function, \fr{quad1}, can be reformulated as follows 
\equl{ \quadraom = P_{XYZ}\langle \Psi^{(1)}(\omega_Y)|\xmz|\Psi^{(1)}(-\omega_Z) \rangle,
}{podst}
where $\xmz = \xmsr$ and $\langle X \rangle = \langle \Psi_0 | X | \Psi_0 \rangle$.
The normalized ground state wave function in the coupled cluster parametrization is given by
\equl{|\Psi_0\rangle = \frac{|\et\Phi\rangle}{\langle \et\Phi|\et\Phi\rangle^{\frac12}}.}{psiz}
The first order 
wave function $\Psi^{(1)}(\omega)$ \rd{ in the coupled cluster parametrization}
is given  by the  operator $\Omega(\omega) = \Omega_1(\omega) + \Omega_2(\omega) + \ldots$, \rd{of the same structure as
the operator $T$},  acting on \rd{$\Psi_0$},\cite{moszynski2005time}
\equl{|\Psi^{(1)}(\omega)\rangle =  (\Omega_0 + \Omega(\omega)) \frac{|\et\Phi\rangle}{\langle \et\Phi|\et\Phi\rangle^{\frac12}},
\qquad \rd{\Omega_0 = - \bb{\Psi_0}{\Omega(\omega)\Psi_0}},}{psi1}
where $\Omega_0$ is a number to ensure the orthogonality of 
$\Psi^{(1)}$ to $\Psi_0$.
\rd{The excitation operator $\Omega(\omega)$ can be found 
  from the following equation}
\cite{moszynski2005time, korona2006time} 
\equl{
\braket{\mu}{\cm{e^{-T} H e^T}{ \Omega(\omega)} + \omega\Omega(\omega)
  + e^{-T}Xe^T} = 0.
}{om}

\rd{We express the excitation operator $\Omega^Y(\omega)$ in the basis of the right eigenvectors $r_N$
of the CC Jacobian matrix $A_{\mu_n\nu_m} = \braket{\widetilde{\mu}_n}{\cm{e^{-T}He^T}{{\nu}_m}}$, using the transformation from the
molecular orbital basis $\mu_n$ to the Jacobian basis $r_N$
\equ{\mu_n = \sum_N \mathcal{L}_{\mu_n N}^{\star}  r_N}
\equsl{
\Omega^Y(\omega) &= \sum_N \sum_{n=1} \sump_{\mu_n}
\mathcal{L}_{\mu_n N}^{\star} O_{\mu_n}^Y(\omega) r_{N} \\
&= \sum_{N} O_{N}^Y(\omega) r_{N}.
}{om-jak}
  \rd{where $\sum_{\mu_n}'$ stands for restricted summation over 
non-redundant  double excitations $ai \ge bj$ and triple
 excitations $ai \ge bj \ge ck$.}
We obtain the amplitudes $O_N^Y(\omega)$ in terms of the right eigenvector $r_N$, by projecting \fr{om} onto the left eigenvector $l_N$ of the Jacobian
\equ{O_N^Y(\omega_Y) = -\frac{\bb{l_N}{\etm Y\et}}{\omega_N + \omega_Y}.\\
}
}

By inserting \fr{psi1} into \fr{podst} we arrive at
\equal{&\rd{\quadraom^{XCC}}  = \nonumber\\
\rd{P_{XYZ}} \Big (&\langle (\omzy + \omy)\pz|\xmz|(\omzz + \omz)\pz\rangle =\nonumber\\
&\brakm{\omy\et}{\et}\brakm{\et}{\omz\et}\frac{\brakett{\et}{\xmz}{\et}}{\mian}\nonumber\\
&-  \brakm{\omy\et}{\et}\frac{\brakett{\et}{\xmz}{\omz \et}}{\mian}\nonumber\\
&- \brakm{\et}{\omz\et}\frac{\brakett{\omy\et}{\xmz}{\et}}{\mian}\nonumber\\
&+ \frac{\brakett{\omy \et}{\xmz}{\omz \et}}{\mian}\Big),
}{wyr}
where $\Omega^V(\omega_V)$ is solution of \fr{om} with $X=V$ and $\omega=\omega_V$.
Further algebraic manipulations are carried out by using the following identities 
\equl{\cm{\et}{\Omega}=0,}{id1}
\equ{\esdm \Phi = \Phi,}
\equ{X\Phi = \langle X\rangle\Phi + \phm(X)\Phi,}
\equl{\frac{\brakett{\et}{X}{\et}}{\mian} = \sred{\esd\etm X \et \esdm},}{id4}
so that the final expression for $\rd{\quadraom^{XCC}}$ reads 
\begin{widetext}
  \equal{&\rd{\quadraom^{XCC}}= \nonumber\\
&\rd{P_{XYZ}}\Big(\bb{(\phm(\esm\etd\omy \etdm\es)}{\esd\etm (\xmz)\et\esdm \phm(\esdm\omz\esd)}\Big).
}{quad}
\end{widetext}

Therefore, by using the eigenvectors and eigenvalues of the CC Jacobian one can express $\rd{\quadraom^{XCC}} $ as follows
\equs{\rd{\quadraom^{XCC}} &= \rd{P_{XYZ}}\sum_{\substack{K=1\\N=1}}\left(O_K^Y(\omega_Y)\right)^{\star}O_N^Z(\rd{-}\omega_Z)
  \bb{\kappa({r_K})}{\esd\etm \xmz\et\esdm \big| \eta({r_N}) }\\
&= \sum_{\substack{K=1\\N=1}}\frac{\bb{\etm Y\et}{l_K}}{\omega_K + \omega_Y}\frac{\bb{l_N}{\etm Z\et}}{\omega_Z - \omega_N}
  \bb{\kappa({r_K})}{\esd\etm \xmz\et\esdm \big| \eta({r_N}) },}
where
\equs{
&\kappa(r_N) = \phm\left(\esm\etd r_N\etdm\es\right),\\
&\eta(r_N) = \phm\left(\esd r_N\esdm\right).
}
Finally, the double residue from the  quadratic response function is given by 
\equsl{\TT_{0L}^Y \TT^X_{LM} \TT^Z_{M0} = &\lim_{\omega_Y\rightarrow -\omega_L}(\omega_L + \omega_Y)\lim_{\omega_Z\rightarrow \omega_\rd{M}}(\omega_\rd{M} - \omega_Z)\quadraom\\
&= \bb{\etm Y\et}{l_L}\bb{\kappa(r_L)}{\esd\etm \xmz\et\esdm  \big| \eta(r_M)}\bb{l_M}{\etm Z\et}.
}{prod}
\rd{We derived our formula for the residue of the quadratic response function,
  so we have to consider the whole right hand side of \fr{prod}. Thus, we cannot
  identify the middle factor on the right hand side of \fr{prod} as $\TT_{LM}^X$. To extract $\TT_{LM}^X$ from \fr{prod},
  we divide both sides by $|\TT_{0L}^Y\TT_{M0}^Z| = \sqrt{|T_{0L}^Y|^2|T_{M0}^Z|^2}$, where
  \equl{|T_{0L}^Y|^2  = \bb{\etm Y\et}{l_L}\bb{\kappa(r_L)}{\eta(r_L)}\bb{l_L}{\etm Y\et}. }{ppp}
  \fr{ppp} is derived by taking double residue of $\langle \Psi^{(1)}(\omega_Y)|X|\Psi^{(1)}(-\omega_Z) \rangle$ with $L=M$ and $Y=Z$. For the
  exact wave function $|T_{0L}^Y|^2 = \brakett{0}{Y}{L}\brakett{L}{Y}{0}$ 
}
This quantity is  then used to extract $\TT_{LM}^X$ from the double residue of the quadratic response function
\equsl{\TT_{LM}^X
  & = \pm \frac{\brakett{0}{Y}{L}\brakett{L}{X_0}{M}\brakett{M}{Z}{0}}
    {\sqrt{\brakett{0}{Y}{L}\braket{L}{L}\brakett{L}{Y}{0}\brakett{0}{Z}{M}\braket{M}{M}\brakett{M}{Z}{0}}}\\
    & = \pm \frac{\lim\limits_{\footnotesize{\omega_Y\rightarrow -\omega_L}}(\omega_L + \omega_Y)\lim\limits_{\omega_z\rightarrow \omega_M}(\omega_M - \omega_Z) \quadraom}
     {\rd{\sqrt{|T_{0L}^Y|^2|T_{M0}^Z|^2}}}.
}{tkn}
The $\pm$ sign results from taking the square root of $|\TT_{0L}^Y|^2$.
This fact is of no concern as both $\TT^X_{LM}$  and $\TT^X_{ML}$
have identical denominators and  we compute the transition strengths which are products $\TT^X_{LM}\TT^X_{ML}$.

The final expression for
$T_{LM}^{X}$ in the XCC theory is given by
\equsl{\TT_{LM}^{X} &= \pm
  \frac{\xi_L^Y\bb{\kappa(r_L)}{\esd\etm \xmz\et\esdm \eta(r_M)}\xi_M^Z}
    {\sqrt{\xi_L^Y \braket{\kappa(r_L)}{\eta(r_L)}(\xi_L^Y)^{\star}\xi_M^Z \braket{\kappa(r_M)}{\eta(r_M)}(\xi_M^Z)^{\star}}}\\
    & = \pm\frac{\bb{\kappa(r_L)}{ \esd\etm \xmz\et\esdm \eta(r_M)}}{\sqrt{\braket{\kappa(r_L)}{\eta(r_L)}\braket{\kappa(r_M)}{\eta(r_M)}}},
}{glowne}
\normalsize
where 
\equ{
\xi_M^Z = \big\langle l_M \big| \etm Z \et \big\rangle.
} 
Note that our formula for $\TT_{LM}^X$ is expressible solely in terms of  commutators. Therefore, it is automatically
size-consistent no matter the level of truncation of the $T$ and $S$ operators.

Alternatively, one can use the identities (\ref{id1}) --- (\ref{id4}) to obtain
\equ{\widetilde{\TT}_{LM}^{X} = \pm\frac{\bb{\eta(r_L)}{ \esm\etd \xmz\etdm\es \kappa(r_M)}}{\sqrt{\braket{\kappa(r_L)}{\eta(r_L)}\braket{\kappa(r_M)}{\eta(r_M)}}}.}
It is easy to notice that as long as
\equl{\TT_{LM}^{X} = \widetilde{\TT}_{LM}^{X}}{tild}
the Hermiticity relation $\TT_{LM}^X = (\TT_{ML}^X)^{\star}$ is satisfied. \fr{tild} is true for any truncated $T$
operator, and the exact $S$ operator. This follows from the fact that
in the derivation of the expression for $\TT_{LM}^X$ we used the definition from \fr{opers} which is valid only for the exact $S$ operator.\cite{jeziorski1993explicitly} Thus,  the Hermiticity relation does not hold for an approximate $S$ operator.
However,
the deviations from the exact symmetry are very small (see \frs{asd}).
\subsection{Approximations}\label{approx}
\rd{In order to obtain computationally tractable expressions for the transition moments we employ several levels of
approximations to \fr{glowne}. There are three issues that we need to address in this equation: the level of truncation
of the operator $T$, operator $S$, and of the multiply nested commutators resulting from the Baker-Campbell-Hausdorff
expansion. We already stated that we employ the operator $T$ from the CCSD/CC3 theory, and that we employ the
approximate operator $S$ defined by \fr{s-approx}.
  To establish the best approximation of the multiply nested commutators we performed the following procedure. We
derived the orbital expressions separately for $S^X_{LM}(m) = (\TT_{LM}^X \TT_{ML}^X)(m)$, $m \in (0, 1, 2, 3, 4)$ where
$m$ is the leading-order in the many-body perturbation theory (MBPT). We computed transition strength
$S^X_{LM}(m)/S^X_{LM}(4)$ for the selected singlet and triplet transitions in the Mg and Sr atoms. In \Frf{mg1a} we plotted the
obtained transition strengths (normalized to $S^X_{LM}(4)$ for more clear view) versus the MBPT order $m$. We studied
the behavior of the numerical values of the transition strength with the increase of the MBPT order and concluded that
in every case the results converge to the numerical limit with the inclusion of third-order terms. Therefore in all our
computations we approximate the XCC transition strength to the third order in MBPT. It should also be mentioned that due
to the computational limits for larger basis sets we discarded terms that scaled as $N^{7}$, with $N$ being a measure of
the system size. We tested that those terms were of negligible importance.
  We want to clearly state here that the only approximation responsible for possible Hermiticity violation in XCC transition strength expression is the truncation of the operator $S$.
  }

\begin{figure}
   \centering  \includegraphics[width=0.7\linewidth]{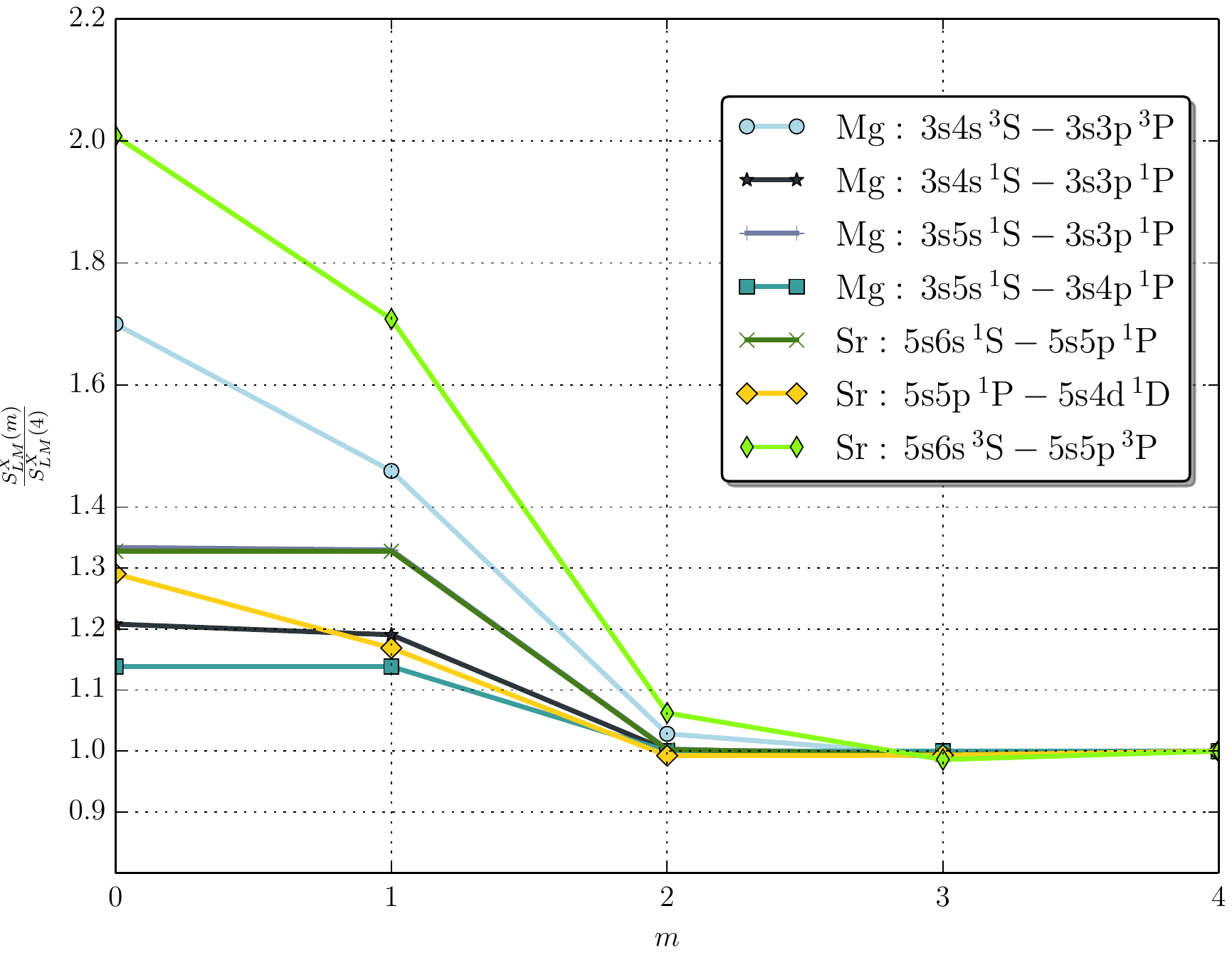}
  \caption{Convergence of the XCC transition strengths with the MBPT order ($m$) for transition dipole strengths for Mg and Sr atoms. The $T$ amplitudes are at the CC3/CCSD level of theory.}
      \label{mg1a}
\end{figure}

\section{Numerical Results}\label{num}
\subsection{Basis sets}

Slater-type orbitals (STOs) used in this work were constructed according to the correlation-consistency principle,\cite{dunning1989gaussian}
similarly as by \citet{lesiukIII} for the beryllium atom. The only difference in the procedure is that the exponents $\zeta$ were chosen according to the well-tempered formula, $(\zeta_{il} = \alpha_l + \beta_li+\gamma_l i^2/n + \delta_l i^3/n^2)$, where $n$ is the number of basis set functions for a
given angular momentum, $l$. After some numerical experimentation, the value of $\delta_l$ was set equal to zero for
$l>2$. A detailed composition of the STOs basis sets is available from the authors upon request. \rd{
STOs basis sets are usually significantly smaller when compared with the Gaussian-type basis sets of a comparable
quality. Therefore there is a strong reason to use them in the computationally demanding coupled cluster theory.}

In \Frt{tab-energy-sing} we demonstrate how the underlying coupled cluster approximation (CCSD/CC3) and
the basis set (Gaussian/Slater) affect the calculated excitation energies for the magnesium atom.
While including  the connected triple amplitudes is important, the use of the Slater-type orbitals (STOs)  yields a dramatic improvement in the accuracy of the excited states energies
\begin{table}[!ht]
    \begin{threeparttable}
    \captionsetup{width=\columnwidth,justification=RaggedRight}
\caption{Singlet and triplet energy levels (cm$^{-1}$) of the magnesium atom computed using Gaussian (G) and Slater (S) basis sets. \rd{$E_{\rm exp}$ is given as an absolute value and the computed energies are given as deviations from the experimental energy.}}
\label{tab-energy-sing}
  \begin{tabular}{l r r r r r r r}
     \toprule
    \text{Level} &
    $E_{\rm exp}$
    & $\text{XCCSD(G)}^{\text{a}} $
    &$\text{XCC3(G)}^{\text{a}}$
    &$ \text{XCCSD(S)}^{\text{b}}$
    & $\mathrm{XCC3(S)}^{\text{b}}$
    &$ \text{XCCSD(S)}^{\text{c}}$
    & $\text{XCC3(S)}^{\text{c}}$\\
    		\midrule
    \ce{3p ^1P^{$\circ$}} & 35051     & -246         & -269      & -13   &  -111    & 69       & -87 \\
    \ce{4s ^1S}          & 43503    & -421         &-413       & -103   & -115     & 37      & -92 \\
    \ce{3d ^1D}          &46403     &   497        & 356       & 194    & -132    & 241     & 121 \\
    \ce{4p ^1P^{$\circ$}} & 49346    &  -394        &-363        & 413   &  443    & 11      & -56 \\
    \ce{5s ^1S}          &52556    &  -214        & -186        & -     &  -      &261     & 168\\
    \ce{3p ^3P^{$\circ$}} & 21891    & -525         &-            &241    &-       & -292    &-\\
    \ce{4s ^3S}          & 41197    & -447        &-            &118     &-       & -110   &-\\
    \ce{4p ^3P^{$\circ$}} & 47848    &-399          &-             &10    &-        & -46   &-\\
    \ce{3d ^3D}          & 47957    & 1325         &-            &-      &-       & -85   & -\\
        \bottomrule
  \end{tabular}
    \begin{tablenotes}
\footnotesize
 \item[a] Gaussian basis set: d-aug-cc-pVQZ\cite{feller1996role, schuchardt2007basis}
 \item[b] Slater basis set: mg-dawtcc4d basis of Lesiuk et al.\cite{lesiukI, lesiukII, lesiukIII}
   with a similar number of basis function as the Gaussian basis set.\\
 \item[c] Slater basis set: mg-dawtcc5d basis of Lesiuk et al.\cite{lesiukI, lesiukII, lesiukIII}
 \end{tablenotes}
  \end{threeparttable}
  \end{table}

    \subsection{Comparison with the QRCC theory}\label{comparison2}
Let us compare our results with the QRCC results obtained with the Dalton program package.\cite{aidas2014dalton}
Although both methods originate from the coupled cluster theory, their working expressions are different and in general,
they are not expected to give identical results.
We computed the first few singlet-singlet transition moments for the Mg atom with both methods. The results are given in
\Frt{comparison-dalton2}. One can see a relatively good agreement between the two methods.

It is clear from \Frt{comparison-dalton2} that the CC3 approximation has a little effect on the transition strength values.
Yet we use the CC3 approximation as it gives better excitation energies, necessary for the lifetime computations.
We also present
the results obtained with the Slater orbitals to emphasize the influence of this basis on
the computed transition strengths. It is worth noting that the use of the Slater orbitals 
 leads in some cases to substantially different results. 

\begin{table*}[!ht]
    \begin{threeparttable}
  \captionsetup{width=\columnwidth,justification=RaggedRight}
\caption{Transition strengths $\SSS_{LM}^X$ (a.u.) in the XCC and QRCC methods for the Mg atom.}
\label{comparison-dalton2}
  \begin{tabular}{l d d d d d d }
\toprule
                \multicolumn{1}{c}{Transition} &
                \multicolumn{1}{c}{XCCSD(G)$^{\text{a}}$} &
                \multicolumn{1}{c}{XCC3(G)$^{\text{a}}$}&
                  \multicolumn{1}{c}{QRCCSD(G)$^{\text{a}}$}&
                    \multicolumn{1}{c}{QRCC3(G)$^{\text{a}}$}&
                    \multicolumn{1}{c}{XCCSD(S)$^{\text{b}}$}&
                    \multicolumn{1}{c}{XCC3(S)$^{\text{b}}$}
                  \\
\midrule
 3s4s \ce{^1S} - 3s3p \ce{^1P^{$\circ$}}    &16.2 & 16.0& 18.3& 18.3& 16.0&15.8\\
  3s4p \ce{^1P^{$\circ$}} - 3s4s \ce{^1S}	& 70.4&  69.9& 73.7& 69.6& 71.6&70.8\\
  3s5s \ce{^1S} - 3s4p \ce{^1P^{$\circ$}}    &  101.8&  101.7& 101.6& 101.6& 97.8&98.2\\
  3s5s \ce{^1S} - 3s3p \ce{^1P^{$\circ$}}	&  0.3&  0.3&0.4 &0.4 &0.3 &0.3\\
 3s3d \ce{^1D} - 3s3p \ce{^1P^{$\circ$}}    &12.7&  12.2& 10.3& 10.0& 23.7& 20.3\\
   3s4p \ce{^1P^{$\circ$}} - 3s3d \ce{^1D}    & 41.8&  42.4& 43.0& - ^{\star}& 86.2& 79.6\\
   \bottomrule
  \end{tabular}
  \begin{tablenotes}
    \footnotesize
    \item[a] Gaussian basis set: d-aug-cc-pVQZ\cite{feller1996role, schuchardt2007basis}
   \item[b] Slater basis set: mg-dawtcc5d basis of Lesiuk et al.\cite{lesiukI, lesiukII, lesiukIII}
      \item [$\star$] Non-physical value. For details see \frs{asd}.
 \end{tablenotes}
  \end{threeparttable}
\end{table*}
\subsection{Comparison with the available theoretical and experimental data}\label{other}
In \Frt{s-chang} we present a comparison of our computed transition strengths with other theoretical approaches, the relativistic multiconfigurational
Hartree Fock approximation (Fischer\cite{fischer1975theoretical}), the CI approximation with the  $B$-spline basis
(Chang and Tang\cite{chang1986effect}), and
the semi-empirical weakest bound electron potential model (Zheng et al.\cite{zheng2001transition}).
The $\SSS_{LM}^X$ values of Chang and Tang  were derived from $A_{LM}^X$ with the experimental excitation energies.
\begin{table}[!ht]
    \begin{threeparttable}
  \captionsetup{width=\columnwidth,justification=RaggedRight}
\caption{Transition strengths $\SSS_{LM}^X$ (a.u.) for the Mg atom.}
\label{s-chang}
  \begin{tabular}{l d d d d d}
\toprule
Transition& \multicolumn{1}{c}{XCC3(G)$^{\text{a}}$} &\multicolumn{1}{c}{XCC3(S)$^{\text{b}}$} & 
\multicolumn{1}{c}{Chang} & \multicolumn{1}{c}{Fischer}& \multicolumn{1}{c}{Zheng}\\
\midrule
  3s4s \ce{^1S} - 3s3p \ce{^1P^{$\circ$}}& 16.0 & 15.8	& 17.9 &18.1 & 18.8 \\
  3s4p \ce{^1P^{$\circ$}} - 3s4s \ce{^1S}&69.9 & 70.8		&69.9		&	65.4 & 77.2\\
  3s5s \ce{^1S} - 3s4p \ce{^1P^{$\circ$}}&	101.8& 98.2		& 91.7		& 92.3	 & 87.4\\
  3s5s \ce{^1S} - 3s3p \ce{^1P^{$\circ$}}&	0.3& 0.3		& 0.4		& 0.3	 & 0.9\\
 3s3d \ce{^1D} - 3s3p \ce{^1P^{$\circ$}} &	12.2& 20.3		&21.5		&21.4	 & 61.5 \\
   3s4p \ce{^1P^{$\circ$}} - 3s3d \ce{^1D}&42.4& 79.6		& 76.6		& 81.9	 & 83.7\\
                \bottomrule
  \end{tabular}
      \begin{tablenotes}
\footnotesize
\item[a] Gaussian basis set: d-aug-cc-pVQZ\cite{feller1996role, schuchardt2007basis}
   \item[b] Slater basis set: mg-dawtcc5d basis of Lesiuk et al.\cite{lesiukI, lesiukII, lesiukIII}
 \end{tablenotes}
\end{threeparttable}
\end{table}

The XCC3(S) results are in a much better agreement with the results calculated with other theoretical methods than the 
results obtained with the XCC3(G) and with QRCC3(G) methods. The most dramatic improvement is observed
for the \ce{3d ^1D - 3p ^1P^{$\circ$}} and  \ce{4p ^1P^{$\circ$} - 3d ^1D} transitions.

The combination of the XCC3 method and the STOs basis set results in lifetimes of the excited states of the Mg atom
in a very good
agreement with the available experimental and theoretical data (\Frttt{lifes-sing}{lifes-trip}).
For the singlet states, we find an excellent agreement with the most recent experimental data,\cite{gratton2003abundances} but not with the
older experiment of Schaefer.\cite{schaefer1971measured} \rd{The mean absolute percentage error of our results for the
singlet states is about 8\% relative to the data of Gratton\cite{gratton2003abundances} and the largest error,
slightly above 10\%, is found for the $\mbox{3s4s}\,^1\mbox{S}$ state.}
Our results are \rd{also} consistent with the lifetimes computed by Froese\cite{fischer1975theoretical} and
Chang,\cite{chang1986effect} but they are in a significant disagreement
with the semi-empirical values of Zheng.\cite{zheng2001transition} \rd{Note parenthetically that no experimental
uncertainty is attributed to some of the values given in \Frttt{lifes-sing}{lifes-trip}, and thus it is difficult to
access their reliability in several cases.}

\begin{table}[!ht]
    \begin{threeparttable}
  \captionsetup{width=\columnwidth,justification=RaggedRight}
  \caption{Lifetimes (in ns) of the singlet excited states of the magnesium atom. Years of publication are given in parentheses.}
  \label{lifes-sing}
  \begin{tabular}{l d d d d d}
\toprule
Ref. & \multicolumn{1}{c}{3s3p \ce{1P^{$\circ$}}} & \multicolumn{1}{c}{3s4s \ce{^1S} }  &
\multicolumn{1}{c}{3s3d \ce{^1D} } & \multicolumn{1}{c}{3s4p \ce{^1P^{$\circ$}} } & \multicolumn{1}{c}{3s5s \ce{^1S} } \\
\midrule
\multicolumn{1}{c}{}&\multicolumn{5}{c}{Experiment}\\
\hline
\text{Gratton(2003)\cite{gratton2003abundances}} & - &46.2\pm2.6 &74.8\pm3& 14.3&101.0\pm3.5\\
\text{Chantepie(1989)\cite{chantepie1989time}} & 2.3 &44.0\pm 5 &72.0\pm4 &13.4\pm0.5 &102.0\pm5 \\
\text{Jonsson(1984)\cite{jonsson1984natural}}& -&47.0\pm3&81.0\pm6&-&100.0\pm5\\
\text{Schaefer(1971)\cite{schaefer1971measured}}& -&-&57.0\pm4&-&163.0\pm8\\
\hline
\multicolumn{1}{c}{}&\multicolumn{5}{c}{Theory}\\
\hline
\text{Fischer(1975)\cite{fischer1975theoretical}} &2.1&44.8&77.2&13.8&102.0\\
\text{Chang(1986)\cite{chang1986effect}}& 2.1&45.8&79.5&14.3&100.0\\
\text{Zheng(2001)\cite{zheng2001transition} }&-&42.3&27.4&-&65.3\\
\text{QRCC3(G)$^{\text{a}}$}&2.1&47.0&200&-^\star&99.8\\
\text{XCC3(G)$^{\text{a}}$}&2.1&53.8&163.9&14.6&91.9\\
\text{XCC3(S)$^{\text{b}}$}&2.1&51.7&79.7&14.1&111.9\\
\bottomrule
  \end{tabular}
 \begin{tablenotes}
\footnotesize
 \item[a] Gaussian basis set: d-aug-cc-pVQZ\cite{feller1996role, schuchardt2007basis}
 \item[b] Slater basis set: mg-dawtcc5d basis of Lesiuk et al.\cite{lesiukI, lesiukII, lesiukIII}
 \item[$\star$] Not converged
 \end{tablenotes}
\end{threeparttable}
\end{table}
All  the computed lifetimes for the triplet states of Mg agree well with the existing experimental and theoretical
results (\Frt{lifes-trip}).
Remarkably, the XCCSD(S) results are close to the most recent experimental
data of Aldenius\cite{aldenius2007experimental} for all states where the data are available. \rd{The mean absolute
percentage deviation from this data is about 8\% and the largest error is found for the $\mbox{3s4s}\,^3\mbox{S}$
state.}
For the 3s5s $^3$S state
other theoretical
results support the older values of Schaefer\cite{schaefer1971measured}
and Gratton.\cite{gratton2003abundances} Similarly, in the case of
the 3s4s $^3$S state, the lifetimes calculated at the XCCSD(S) level are slightly larger than the other
theoretical results, yet in an excellent agreement with the Aldenius
experiment.\cite{aldenius2007experimental} For the 3s4p $^3$P state there are no experimental results available,
but all the theoretical lifetimes, including the XCCSD(S) one, are consistent within 10\% at worst. 
The triplet-triplet transition dipole moments which are necessary to compute the
lifetimes of the triplet states are not available in the QRCC 
implementation. Therefore, no comparison with the QRCC method is possible.

\begin{table}[!ht]
    \begin{threeparttable}
  \captionsetup{width=\columnwidth,justification=RaggedRight}
  \caption{Lifetimes (in ns) of the triplet excited states for the Mg atom. Years of publication are given in parentheses.}
\label{lifes-trip}
  \begin{tabular}{l d d d d}
\toprule
Ref & \multicolumn{1}{c}{3s4s \ce{^3S}} & \multicolumn{1}{c}{3s5s \ce{^3S} }  &
\multicolumn{1}{c}{3s4p \ce{^3P} } & \multicolumn{1}{c}{3s3d \ce{^3D} }\\
\midrule
\multicolumn{1}{c}{}&\multicolumn{4}{c}{Experiment}\\
\hline
\text{Aldenius(2007)\cite{aldenius2007experimental}} & 11.5\pm 1.0& 29.0 \pm 0.3& -& 5.9\pm 0.4\\
\text{Kwiatkowski(1980)\cite{kwiatkowski1980lifetime} }& 9.7\pm 0.6& - & -& 5.9\pm 0.4\\
\text{Andersen(1972)\cite{andersen1972lifetimes}} & 10.1\pm 0.8& - & -& 6.6\pm 0.5\\
\text{Schaefer(1971)\cite{schaefer1971measured}}  & 14.8\pm0.7&25.6 \pm 2.1&- &11.3\pm 0.8 \\
\text{Ueda(1982)\cite{ueda1982measurements}}  &9.9 \pm 1.25 & - & -&5.93 \\
\text{Havey(1977)\cite{havey1977direct}} &9.7\pm 0.5 & - &- & -\\

\text{Gratton(2003)\cite{gratton2003abundances}} & 9.8 \pm 0.3&  25.6 \pm 2.1& -& -\\
\hline
\multicolumn{1}{c}{}&\multicolumn{4}{c}{Theory}\\
\hline
\text{Fischer(1975)\cite{fischer1975theoretical}} & 9.86& 26.8 & 74.5&6.18 \\
\text{Moccia(1988)\cite{moccia1988atomic}} & 9.7& 26.5 &81.0 &5.8 \\
\text{Victor(1976)\cite{victor1976oscillator}}& 9.07&-  &- & 6.25\\
\text{Chang(1986)\cite{chang1986effect}}& 9.98& 27.5 &77.0 &5.89 \\
\text{Mendoza(1981)\cite{mendoza1981term}}& 9.79& - & -&- \\
\text{Zheng(2001)\cite{zheng2001transition} }&- &-  &78.49 &- \\
\text{XCCSD(S)$^{\text{a}}$}&12.7&29.87&70.44&5.33\\
\bottomrule
  \end{tabular}
  \begin{tablenotes}
  \footnotesize
   \item[a] Slater basis set: mg-dawtcc5d basis of Lesiuk et al.\cite{lesiukI, lesiukII, lesiukIII}
 \end{tablenotes}
\end{threeparttable}
\end{table}

In  \Frt{lifesr} we present transition probabilities for the Sr atom. For the singlet states we note a good agreement with the
Werij\cite{werij1992oscillator} results.  For the 5s5p\ce{^1P^{$\circ$}}-5s4d\ce{^1D} transition our result is also within the
experimental error of \citet{hunter1986observation}.
In the case of 5s6s\ce{^3S}-5s5p\ce{^3P^{$\circ$}} transition, our result deviates significantly from other theoretical and experimental
results. The 5s4d\ce{^3D}-5s5p\ce{^3P^{$\circ$}} transition strengths vary between different theories and experiments to a large degree. Our result
is in reasonable agreement with the latest theoretical result of \citet{porsev2008determination}. 
\begin{table}[!ht]
    \begin{threeparttable}
  \captionsetup{width=\columnwidth,justification=RaggedRight}
\caption{Transition probabilities ($10^6\, \mbox{s}^{-1}$) of the Sr atom.}
\label{lifesr}
\begin{tabular}{l d d d d}
\toprule
Ref & \multicolumn{1}{c}{5s6s\ce{^1S}-5s5p\ce{^1P^{$\circ$}}} & \multicolumn{1}{c}{5s5p\ce{^1P^{$\circ$}}-5s4d\ce{^1D}}  &
\multicolumn{1}{c}{5s6s\ce{^3S}-5s5p\ce{^3P^{$\circ$}}} & \multicolumn{1}{c}{5s4d\ce{^3D}-5s5p\ce{^3P^{$\circ$}}}\\
\midrule
\multicolumn{1}{c}{}&\multicolumn{4}{c}{Experiment}\\
\hline
\text{Hunter(1986)\cite{hunter1986observation}} &-  &  0.0039\pm0.0016& - &-  \\
\text{Jonsson(1984)\cite{jonsson1984natural}} &-  &-  &66.0 \pm 4 &-  \\
\text{Brinkmann(1969)\cite{brinkmann1969experimente}} &-  & - &  91.0 \pm 2.5&-  \\
\text{Havey(1977)\cite{havey1977direct}} & - & - & 77.0 \pm 4.5 & - \\
\text{Borisov(1987)\cite{borisov87}} & - &-  &-  &  0.24\pm 0.04\\
\text{Miller(1992)\cite{miller1992collisional}} & - & - & - &  0.29 \pm 0.03\\
\hline
\multicolumn{1}{c}{}&\multicolumn{4}{c}{Theory}\\
\hline
\text{Werij(1992)\cite{werij1992oscillator}} & 18.6 & 0.0017&71.3  &  4.32\\
\text{Vaeck(1988)\cite{vaeck1988multiconfiguration}} &  -& 0.0048 &-  &-  \\
\text{Porsev(2008)\cite{porsev2008determination}} &-  &-  &  70.9& 0.41 \\
\text{XCC3(G)$^{\text{a}}$} & 15.1 &  0.0027&  47&  0.70\\
\text{QRCC3(G)$^{\text{a}}$} & 20.4 & -^{\circ} &-^{\star}  &-^{\star}  \\
\bottomrule
\end{tabular}
    \begin{tablenotes}
\footnotesize
\item[a] Gaussian basis set: [8s8p5d4f1g] basis augmented by a set of [1s1p1d1f3g] diffuse functions and the ECP28MDF pseudopotential\cite{feller1996role, schuchardt2007basis, skomorowski2012rovibrational, lim2006relativistic}
  \item[$\circ$] Not converged
\item[$\star$] Not implemented
 \end{tablenotes}
\end{threeparttable}
\end{table}

\subsection{Possible Hermiticity violation and its consequences}\label{asd}

The exact transition moment $\TT_{LM}^X$ is Hermitian, i.e., it satisfies the relation given by \fr{hermit0}.
This implies that the transition strength $\SSS_{LM}^X$, \fr{ts}, cannot be negative.
This condition is not satisfied in the QRCC theory as well as in the approximate XCC theory.
However, in the XCC theory this violation of the Hermiticity originates
solely from the truncation of the $S$ operator, while in the QRCC method it has a more fundamental origin.
Therefore, the lack of the Hermiticity is expected to be a fairly minor issue in our method, by contrast to
the QRCC theory.

For the purpose of this study we investigate some problematic transitions in the Mg atom and \ce{Mg2} molecule which have been
encountered beforehand.\cite{amaran2013femtosecond}
We found that 
the transitions strengths for the \ce{3d ^1D - 3p ^1P^{$\circ$}}, \ce{3d
^1D - 4p ^1P^{$\circ$}}
and \ce{3d ^1D - 5p ^1P^{$\circ$}} transitions computed with the QRCC code exhibited a non-physical behavior, i.e., some of the contributions were negative.
No such artifacts were found in any transition strengths contributions with the XCC theory.
In \Frt{negat} we present the differences between $\TT_{LM}^X$ and $(\TT_{ML}^X)^{\star}$ computed with the QRCC and XCC theories.
In QRCC  these differences are significant, especially in
situations where one is positive and the other is negative. Although in the XCC method the Hermiticity is also violated,
we do not observe such strong deviations.
\begin{table}[!ht]
  \begin{threeparttable}
  \captionsetup{width=\columnwidth,justification=RaggedRight}
\caption{$\TT_{LM}^X$ and $(\TT_{ML}^X)^{\star}$ computed with the QRCC and XCC methods for the Mg atom.}
\label{negat}
\begin{tabular}{l d d d d}
    \toprule
    \multicolumn{1}{c}{Transition}  
    &   \multicolumn{1}{c}{$\TT_{LM}^X(\mathrm{QRCC})$} 
    &   \multicolumn{1}{c}{$(\TT_{ML}^X)^\star(\mathrm{QRCC})$}
            &   \multicolumn{1}{c}{$ \TT_{LM}^X(\mathrm{XCC})$ }
            &   \multicolumn{1}{c}{$(\TT_{ML}^X)^\star(\mathrm{XCC})$}          \\
    \midrule
    \multicolumn{5}{c}{aug-cc-pVQZ}\\
    \midrule
 3s4s \ce{^1S} - 3s3p \ce{^1P^{$\circ$}} & 4.3 & 4.26& 4.00&4.01\\
 3s4p \ce{^1P^{$\circ$}} - 3s4s \ce{^1S}& 8.39 & 8.30 & 8.36 & 8.36 \\
     \midrule
 \multicolumn{5}{c}{d-aug-cc-pVQZ}\\
            \midrule
  3s5s \ce{^1S} - 3s4p \ce{^1P^{$\circ$}}& 10.12 & 10.04 & 10.08 & 10.09 \\ 
  3s5s \ce{^1S} - 3s3p \ce{^1P^{$\circ$}}& 0.60 & 0.60 & 0.51 & 0.51 \\
 3s3d \ce{^1D} - 3s3p \ce{^1P^{$\circ$}} & 0.67& -0.40 & 1.40 & 1.43 \\
   3s4p \ce{^1P^{$\circ$}} - 3s3d \ce{^1D}&-1.18 & 0.72  & 2.64 & 2.63 \\
    \bottomrule
  \end{tabular}
  \end{threeparttable}
\end{table}

A different problem is found for the \ce{Mg2} molecule.  In \Frf{mg2-krzywe-ccsd}
 we present potential energy curves for $(1)^1\Pi_u$, $(2)^1\Pi_u$ and $(1)^1\Sigma_g^+$
 states of \ce{Mg2} computed with the EOM-CCSD approximation.
 \begin{figure}[!ht]
 \includegraphics[width=0.7\textwidth]{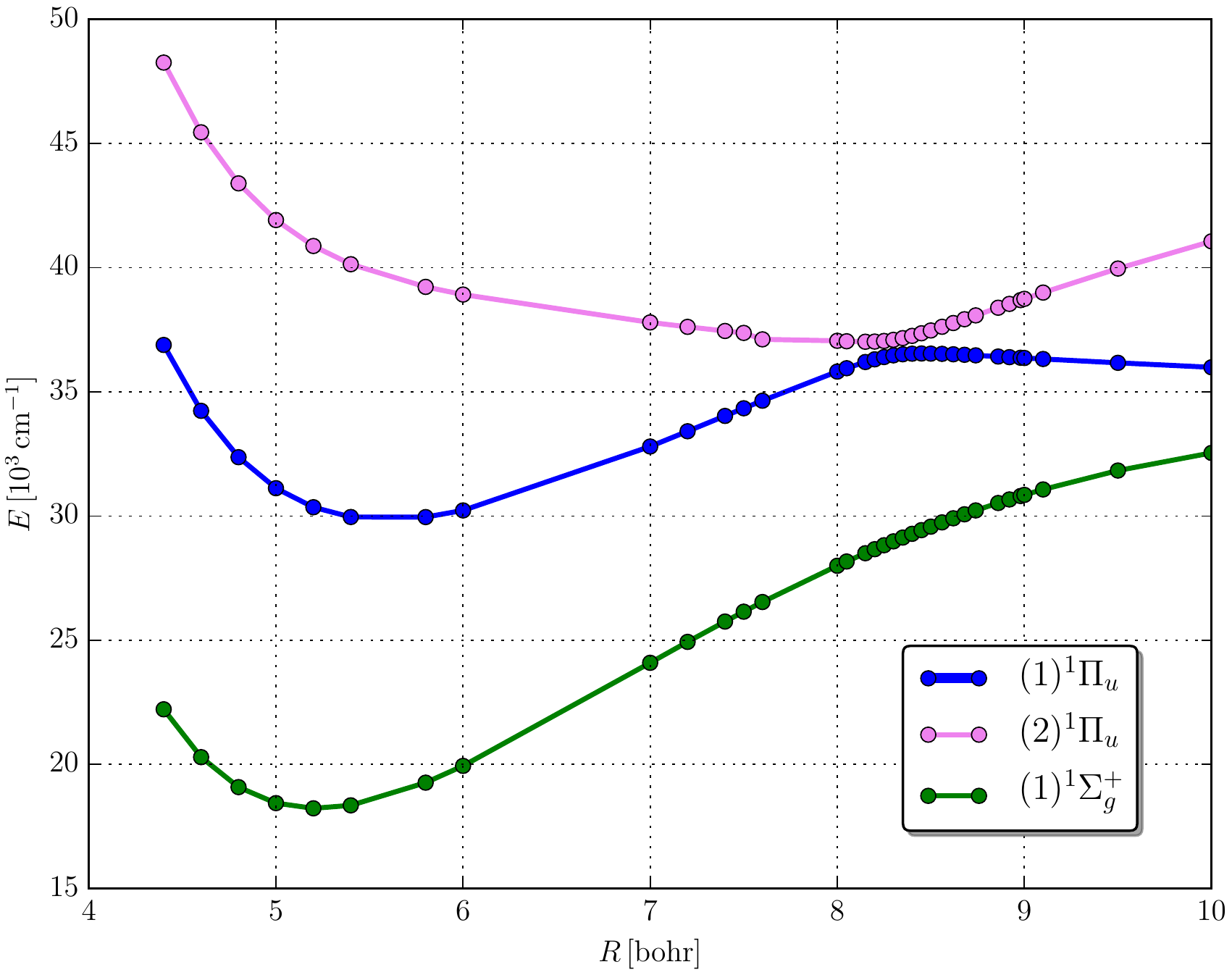}
\caption{Potential energy curves for Mg$_2$ states}
\label{mg2-krzywe-ccsd}
\end{figure}
We also present a set of transition strengths for various interatomic distances $R$ computed
with the XCCSD(G) and QRCCSD(G) methods, \Frf{both}. For $R$ ranging from $7$ to $9$ Bohr both methods give similar results.
\begin{figure}[!ht]
 \includegraphics[width=0.7\textwidth]{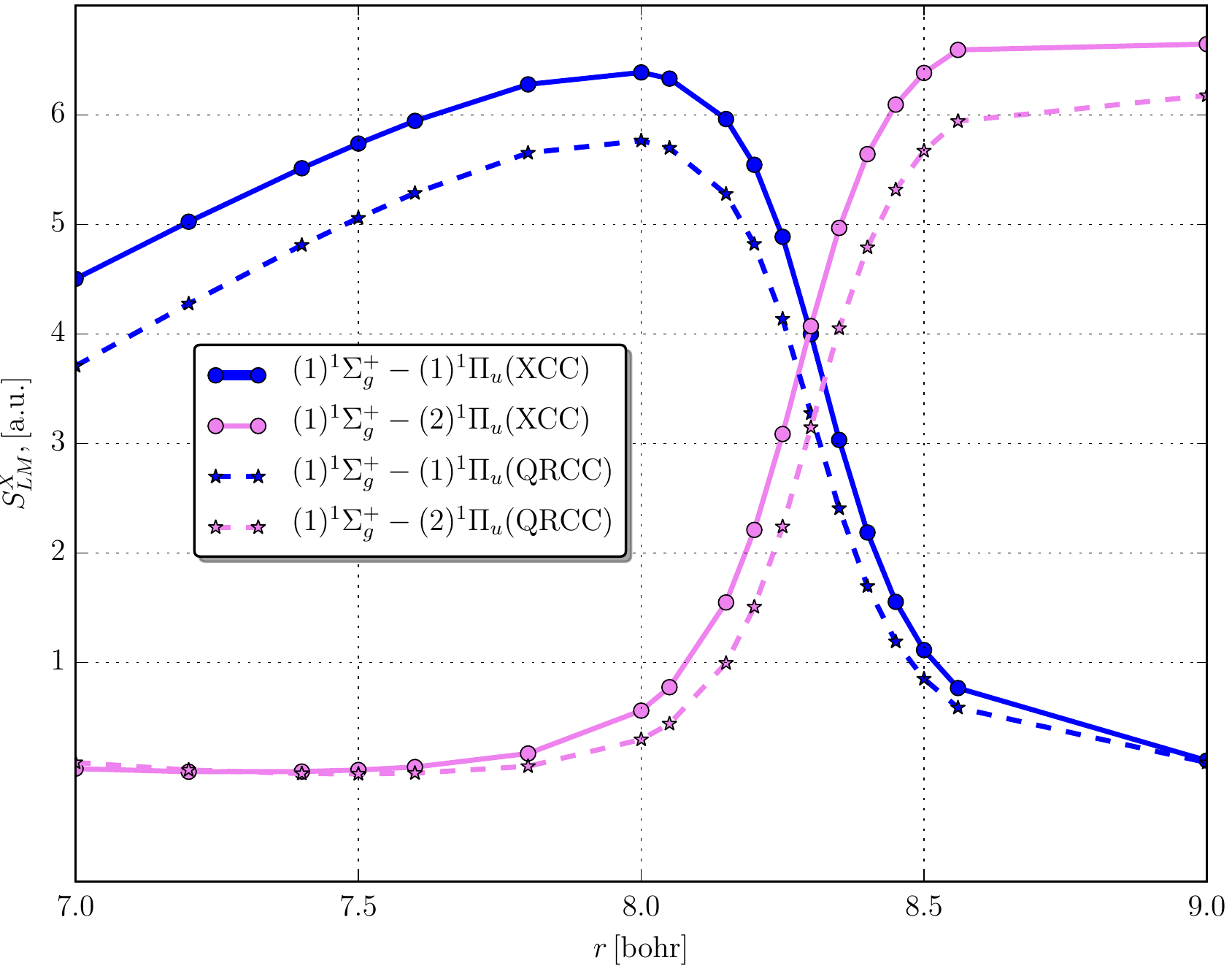}
\caption{Transition strengths for \ce{Mg2} computed with XCCSD(G) and QRCCSD(G) method for $R$ = 7-9 a.u.}
\label{both}
\end{figure}
However, the QRCCSD(G) method exhibits problems at small distances where we obtained negative transition strengths that by definition (\ref{ts}) should always be positive.
In \Frf{mg2-ts-dalton} we see a pole-like structure which is clearly an artifact, as no such structure should be observed
for the transition strengths. By contrast, no such difficulties were found in the XCCSD(G) theory,
see \Frf{mg2-ts-a}.
This suggests that the adopted truncation scheme for the $S$ operator has a negligible impact on the behavior of the XCC
transition moments.
\begin{figure}[!ht]
 \includegraphics[width=0.7\textwidth]{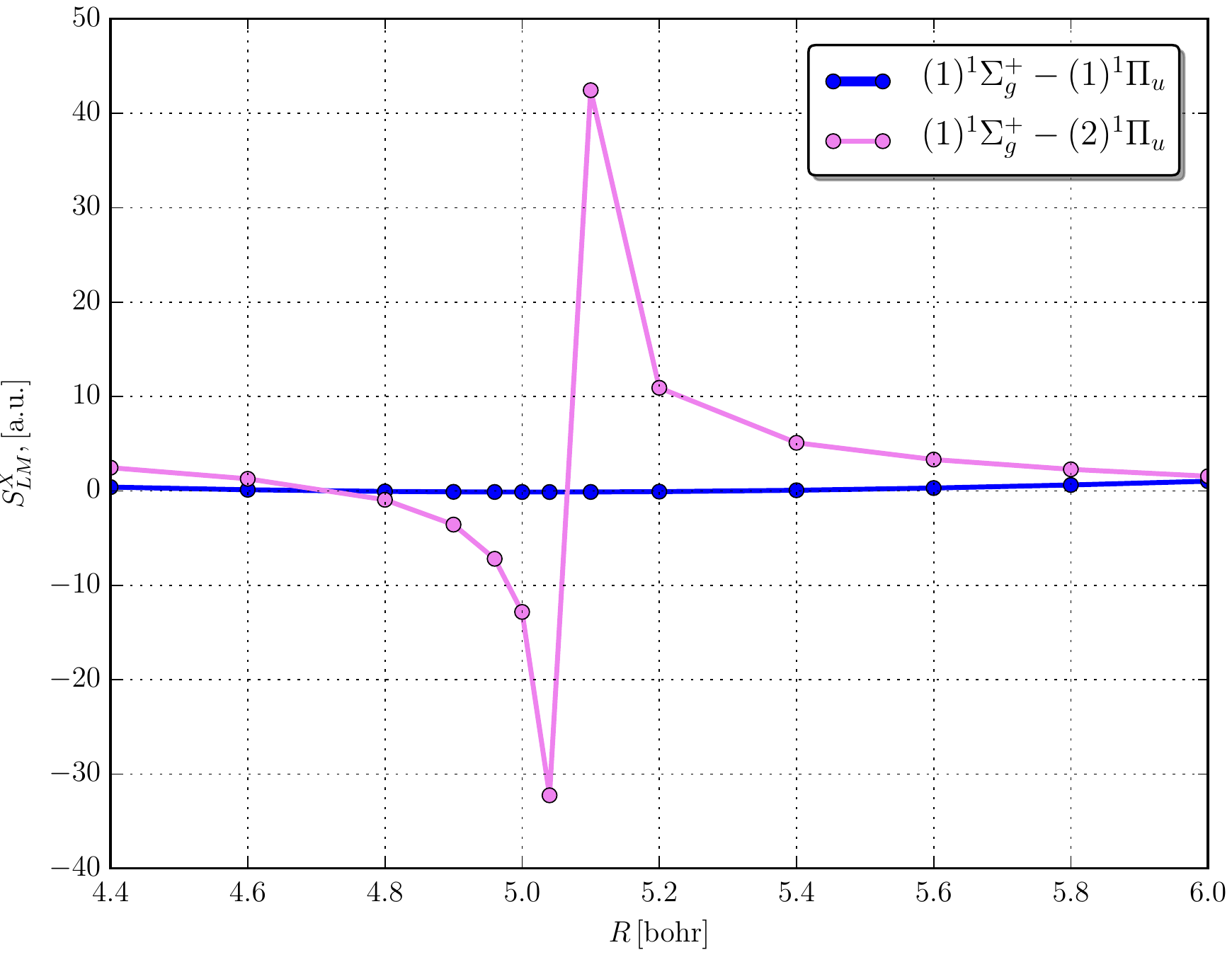}
\caption{Transition strengths for \ce{Mg2} computed with QRCC3(G) method}
\label{mg2-ts-dalton}
\end{figure}
\begin{figure}[!ht]
 \includegraphics[width=0.7\textwidth]{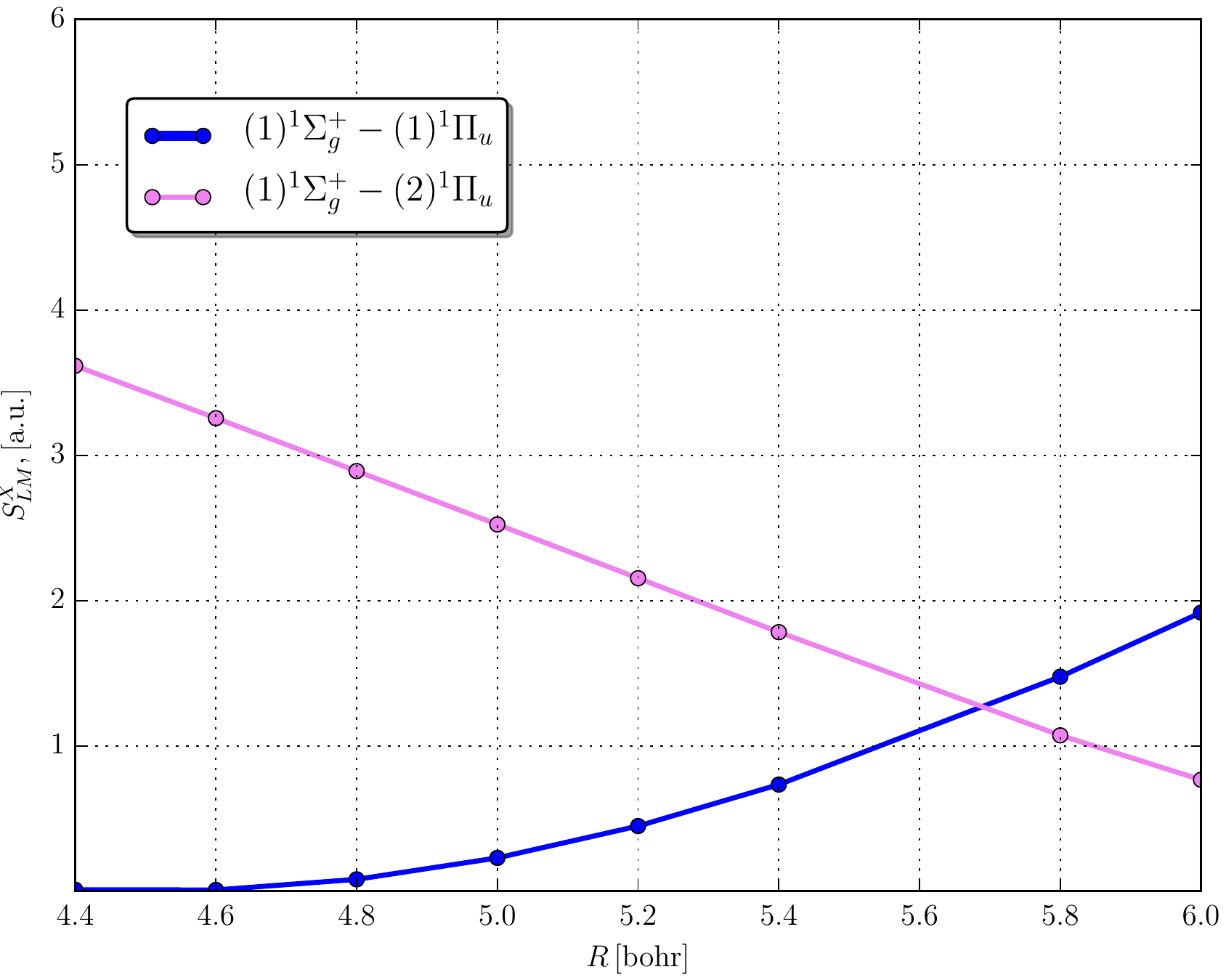}
\caption{Transition strengths for \ce{Mg2} computed with XCCSD method}
\label{mg2-ts-a}
\end{figure}

\section{Conclusions}\label{concl}
We have presented a novel coupled cluster approach to the computation of the transition moments between the excited
electronic states.
In contrast to the existing CC approaches, our method approximately obeys the Hermiticity relation
$\TT_{LM}^X = (\TT_{ML}^X)^{\star}$ and the deviations from this symmetry are negligible. There are three levels of approximations in our formulas
for $\TT_{LM}^X$:\begin{enumerate}
\item the underlying model for the CC amplitudes (CCSD/CC3)
\item approximations of the auxiliary operator $S$ employed in the computation of the expectation values
  with the CC ground state wave function
\item choice of the commutators included in the expansion of the XCC formula for $\TT_{LM}^X$.
\end{enumerate}
In trouble-free situations, i.e., when the existing QRCC approach satisfies the Hermiticity relation to a good
approximation, both methods yield transition moments of a similar quality. However, in certain cases
the QRCC method violates the Hermiticity relation to an unacceptable degree and
gives unphysical  values of the transition strengths. The XCC method does not suffer from this problem. Clearly,
this can be viewed as an important improvement over the existing QRCC approach.

We have presented numerical examples for several singlet-singlet and triplet-triplet dipole transitions in the Mg and Sr
atoms, and the \ce{Mg2} molecule.
Lifetimes derived from the transition moments computed with our method are, in most cases, very close to the
available experimental data and to other theoretical results.
We have assessed the performance of our method in the STOs basis set and obtained results of significantly better
quality than with the available Gaussian basis sets. In certain cases, the use of STOs basis set was the game-changer. 

In two the forthcoming papers we will consider calculations of the radial and angular nonadiabatic coupling matrix elements and
of the spin-orbit coupling matrix elements between the excited states within the XCC theory. Both works are in preparation.

The code for transition moments between the excited states will be incorporated in the
KOŁOS: A general purpose ab initio program for the electronic structure calculation with
Slater orbitals, Slater geminals, and Kołos-Wolniewicz functions.

\section{Acknowledgement}
This research was supported in part by the National Science Foundation under Grant No. NSF PHY-1125915 and
by the National Science Centre (NCN) under Grant No. 2016/21/N/ST4/03734.

\bibliography{ref}

\begin{thebibliography}{57}%
\makeatletter
\providecommand \@ifxundefined [1]{%
 \@ifx{#1\undefined}
}%
\providecommand \@ifnum [1]{%
 \ifnum #1\expandafter \@firstoftwo
 \else \expandafter \@secondoftwo
 \fi
}%
\providecommand \@ifx [1]{%
 \ifx #1\expandafter \@firstoftwo
 \else \expandafter \@secondoftwo
 \fi
}%
\providecommand \natexlab [1]{#1}%
\providecommand \enquote  [1]{``#1''}%
\providecommand \bibnamefont  [1]{#1}%
\providecommand \bibfnamefont [1]{#1}%
\providecommand \citenamefont [1]{#1}%
\providecommand \href@noop [0]{\@secondoftwo}%
\providecommand \href [0]{\begingroup \@sanitize@url \@href}%
\providecommand \@href[1]{\@@startlink{#1}\@@href}%
\providecommand \@@href[1]{\endgroup#1\@@endlink}%
\providecommand \@sanitize@url [0]{\catcode `\\12\catcode `\$12\catcode
  `\&12\catcode `\#12\catcode `\^12\catcode `\_12\catcode `\%12\relax}%
\providecommand \@@startlink[1]{}%
\providecommand \@@endlink[0]{}%
\providecommand \url  [0]{\begingroup\@sanitize@url \@url }%
\providecommand \@url [1]{\endgroup\@href {#1}{\urlprefix }}%
\providecommand \urlprefix  [0]{URL }%
\providecommand \Eprint [0]{\href }%
\providecommand \doibase [0]{http://dx.doi.org/}%
\providecommand \selectlanguage [0]{\@gobble}%
\providecommand \bibinfo  [0]{\@secondoftwo}%
\providecommand \bibfield  [0]{\@secondoftwo}%
\providecommand \translation [1]{[#1]}%
\providecommand \BibitemOpen [0]{}%
\providecommand \bibitemStop [0]{}%
\providecommand \bibitemNoStop [0]{.\EOS\space}%
\providecommand \EOS [0]{\spacefactor3000\relax}%
\providecommand \BibitemShut  [1]{\csname bibitem#1\endcsname}%
\let\auto@bib@innerbib\@empty
\bibitem [{\citenamefont {Zubarev}(1960)}]{zubarev1974nonlinear}%
  \BibitemOpen
  \bibfield  {author} {\bibinfo {author} {\bibfnamefont {D.}~\bibnamefont
  {Zubarev}},\ }\href@noop {} {\bibfield  {journal} {\bibinfo  {journal} {Usp.
  Fiz. Nauk}\ }\textbf {\bibinfo {volume} {3}},\ \bibinfo {pages} {320}
  (\bibinfo {year} {1960})}\BibitemShut {NoStop}%
\bibitem [{\citenamefont {Linderberg}\ and\ \citenamefont
  {{\"O}hrn}(2004)}]{linderberg2004propagators}%
  \BibitemOpen
  \bibfield  {author} {\bibinfo {author} {\bibfnamefont {J.}~\bibnamefont
  {Linderberg}}\ and\ \bibinfo {author} {\bibfnamefont {Y.}~\bibnamefont
  {{\"O}hrn}},\ }\href@noop {} {\emph {\bibinfo {title} {Propagators in quantum
  chemistry}}}\ (\bibinfo  {publisher} {John Wiley \& Sons},\ \bibinfo {year}
  {2004})\BibitemShut {NoStop}%
\bibitem [{\citenamefont {Oddershede}(1987)}]{oddershede1987propagator}%
  \BibitemOpen
  \bibfield  {author} {\bibinfo {author} {\bibfnamefont {J.}~\bibnamefont
  {Oddershede}},\ }\href {\doibase 10.1002/9780470142943.ch3} {\bibfield
  {journal} {\bibinfo  {journal} {Adv. Chem. Phys.}\ }\textbf {\bibinfo
  {volume} {69}},\ \bibinfo {pages} {201} (\bibinfo {year} {1987})}\BibitemShut
  {NoStop}%
\bibitem [{\citenamefont {Rybak}\ \emph
  {et~al.}(2011{\natexlab{a}})\citenamefont {Rybak}, \citenamefont {Amitay},
  \citenamefont {Amaran}, \citenamefont {Kosloff}, \citenamefont {Tomza},
  \citenamefont {Moszynski},\ and\ \citenamefont
  {Koch}}]{rybak2011femtosecond}%
  \BibitemOpen
  \bibfield  {author} {\bibinfo {author} {\bibfnamefont {L.}~\bibnamefont
  {Rybak}}, \bibinfo {author} {\bibfnamefont {Z.}~\bibnamefont {Amitay}},
  \bibinfo {author} {\bibfnamefont {S.}~\bibnamefont {Amaran}}, \bibinfo
  {author} {\bibfnamefont {R.}~\bibnamefont {Kosloff}}, \bibinfo {author}
  {\bibfnamefont {M.}~\bibnamefont {Tomza}}, \bibinfo {author} {\bibfnamefont
  {R.}~\bibnamefont {Moszynski}}, \ and\ \bibinfo {author} {\bibfnamefont
  {C.~P.}\ \bibnamefont {Koch}},\ }\href@noop {} {\bibfield  {journal}
  {\bibinfo  {journal} {Farad. Discuss.}\ }\textbf {\bibinfo {volume} {153}},\
  \bibinfo {pages} {383} (\bibinfo {year} {2011}{\natexlab{a}})}\BibitemShut
  {NoStop}%
\bibitem [{\citenamefont {Rybak}\ \emph
  {et~al.}(2011{\natexlab{b}})\citenamefont {Rybak}, \citenamefont {Amaran},
  \citenamefont {Levin}, \citenamefont {Tomza}, \citenamefont {Moszynski},
  \citenamefont {Kosloff}, \citenamefont {Koch},\ and\ \citenamefont
  {Amitay}}]{rybak2011generating}%
  \BibitemOpen
  \bibfield  {author} {\bibinfo {author} {\bibfnamefont {L.}~\bibnamefont
  {Rybak}}, \bibinfo {author} {\bibfnamefont {S.}~\bibnamefont {Amaran}},
  \bibinfo {author} {\bibfnamefont {L.}~\bibnamefont {Levin}}, \bibinfo
  {author} {\bibfnamefont {M.}~\bibnamefont {Tomza}}, \bibinfo {author}
  {\bibfnamefont {R.}~\bibnamefont {Moszynski}}, \bibinfo {author}
  {\bibfnamefont {R.}~\bibnamefont {Kosloff}}, \bibinfo {author} {\bibfnamefont
  {C.~P.}\ \bibnamefont {Koch}}, \ and\ \bibinfo {author} {\bibfnamefont
  {Z.}~\bibnamefont {Amitay}},\ }\href@noop {} {\bibfield  {journal} {\bibinfo
  {journal} {Phys. Rev. Lett.}\ }\textbf {\bibinfo {volume} {107}},\ \bibinfo
  {pages} {273001} (\bibinfo {year} {2011}{\natexlab{b}})}\BibitemShut
  {NoStop}%
\bibitem [{\citenamefont {Amaran}\ \emph {et~al.}(2013)\citenamefont {Amaran},
  \citenamefont {Kosloff}, \citenamefont {Tomza}, \citenamefont {Skomorowski},
  \citenamefont {Paw{\l}owski}, \citenamefont {Moszynski}, \citenamefont
  {Rybak}, \citenamefont {Levin}, \citenamefont {Amitay}, \citenamefont
  {Berglund}, \citenamefont {Reich},\ and\ \citenamefont
  {Koch}}]{amaran2013femtosecond}%
  \BibitemOpen
  \bibfield  {author} {\bibinfo {author} {\bibfnamefont {S.}~\bibnamefont
  {Amaran}}, \bibinfo {author} {\bibfnamefont {R.}~\bibnamefont {Kosloff}},
  \bibinfo {author} {\bibfnamefont {M.}~\bibnamefont {Tomza}}, \bibinfo
  {author} {\bibfnamefont {W.}~\bibnamefont {Skomorowski}}, \bibinfo {author}
  {\bibfnamefont {F.}~\bibnamefont {Paw{\l}owski}}, \bibinfo {author}
  {\bibfnamefont {R.}~\bibnamefont {Moszynski}}, \bibinfo {author}
  {\bibfnamefont {L.}~\bibnamefont {Rybak}}, \bibinfo {author} {\bibfnamefont
  {L.}~\bibnamefont {Levin}}, \bibinfo {author} {\bibfnamefont
  {Z.}~\bibnamefont {Amitay}}, \bibinfo {author} {\bibfnamefont {J.~M.}\
  \bibnamefont {Berglund}}, \bibinfo {author} {\bibfnamefont {D.~M.}\
  \bibnamefont {Reich}}, \ and\ \bibinfo {author} {\bibfnamefont {C.~P.}\
  \bibnamefont {Koch}},\ }\href@noop {} {\bibfield  {journal} {\bibinfo
  {journal} {J. Chem. Phys.}\ }\textbf {\bibinfo {volume} {139}},\ \bibinfo
  {pages} {164124} (\bibinfo {year} {2013})}\BibitemShut {NoStop}%
\bibitem [{\citenamefont {Levin}\ \emph {et~al.}(2015)\citenamefont {Levin},
  \citenamefont {Skomorowski}, \citenamefont {Rybak}, \citenamefont {Kosloff},
  \citenamefont {Koch},\ and\ \citenamefont {Amitay}}]{levin2015coherent}%
  \BibitemOpen
  \bibfield  {author} {\bibinfo {author} {\bibfnamefont {L.}~\bibnamefont
  {Levin}}, \bibinfo {author} {\bibfnamefont {W.}~\bibnamefont {Skomorowski}},
  \bibinfo {author} {\bibfnamefont {L.}~\bibnamefont {Rybak}}, \bibinfo
  {author} {\bibfnamefont {R.}~\bibnamefont {Kosloff}}, \bibinfo {author}
  {\bibfnamefont {C.~P.}\ \bibnamefont {Koch}}, \ and\ \bibinfo {author}
  {\bibfnamefont {Z.}~\bibnamefont {Amitay}},\ }\href@noop {} {\bibfield
  {journal} {\bibinfo  {journal} {Phys. Rev. Lett.}\ }\textbf {\bibinfo
  {volume} {114}},\ \bibinfo {pages} {233003} (\bibinfo {year}
  {2015})}\BibitemShut {NoStop}%
\bibitem [{\citenamefont {Skomorowski}\ \emph {et~al.}(2012)\citenamefont
  {Skomorowski}, \citenamefont {Paw{\l}owski}, \citenamefont {Koch},\ and\
  \citenamefont {Moszynski}}]{skomorowski2012rovibrational}%
  \BibitemOpen
  \bibfield  {author} {\bibinfo {author} {\bibfnamefont {W.}~\bibnamefont
  {Skomorowski}}, \bibinfo {author} {\bibfnamefont {F.}~\bibnamefont
  {Paw{\l}owski}}, \bibinfo {author} {\bibfnamefont {C.~P.}\ \bibnamefont
  {Koch}}, \ and\ \bibinfo {author} {\bibfnamefont {R.}~\bibnamefont
  {Moszynski}},\ }\href@noop {} {\bibfield  {journal} {\bibinfo  {journal} {J.
  Chem. Phys.}\ }\textbf {\bibinfo {volume} {136}},\ \bibinfo {pages} {194306}
  (\bibinfo {year} {2012})}\BibitemShut {NoStop}%
\bibitem [{\citenamefont {McGuyer}\ \emph
  {et~al.}(2015{\natexlab{a}})\citenamefont {McGuyer}, \citenamefont
  {McDonald}, \citenamefont {Iwata}, \citenamefont {Tarallo}, \citenamefont
  {Skomorowski}, \citenamefont {Moszynski},\ and\ \citenamefont
  {Zelevinsky}}]{mcguyer2015precise}%
  \BibitemOpen
  \bibfield  {author} {\bibinfo {author} {\bibfnamefont {B.}~\bibnamefont
  {McGuyer}}, \bibinfo {author} {\bibfnamefont {M.}~\bibnamefont {McDonald}},
  \bibinfo {author} {\bibfnamefont {G.}~\bibnamefont {Iwata}}, \bibinfo
  {author} {\bibfnamefont {M.}~\bibnamefont {Tarallo}}, \bibinfo {author}
  {\bibfnamefont {W.}~\bibnamefont {Skomorowski}}, \bibinfo {author}
  {\bibfnamefont {R.}~\bibnamefont {Moszynski}}, \ and\ \bibinfo {author}
  {\bibfnamefont {T.}~\bibnamefont {Zelevinsky}},\ }\href@noop {} {\bibfield
  {journal} {\bibinfo  {journal} {Nature Phys.}\ }\textbf {\bibinfo {volume}
  {11}},\ \bibinfo {pages} {32} (\bibinfo {year}
  {2015}{\natexlab{a}})}\BibitemShut {NoStop}%
\bibitem [{\citenamefont {McDonald}\ \emph {et~al.}(2016)\citenamefont
  {McDonald}, \citenamefont {McGuyer}, \citenamefont {Apfelbeck}, \citenamefont
  {Lee}, \citenamefont {Majewska}, \citenamefont {Moszynski},\ and\
  \citenamefont {Zelevinsky}}]{mcdonald2016photodissociation}%
  \BibitemOpen
  \bibfield  {author} {\bibinfo {author} {\bibfnamefont {M.}~\bibnamefont
  {McDonald}}, \bibinfo {author} {\bibfnamefont {B.}~\bibnamefont {McGuyer}},
  \bibinfo {author} {\bibfnamefont {F.}~\bibnamefont {Apfelbeck}}, \bibinfo
  {author} {\bibfnamefont {C.-H.}\ \bibnamefont {Lee}}, \bibinfo {author}
  {\bibfnamefont {I.}~\bibnamefont {Majewska}}, \bibinfo {author}
  {\bibfnamefont {R.}~\bibnamefont {Moszynski}}, \ and\ \bibinfo {author}
  {\bibfnamefont {T.}~\bibnamefont {Zelevinsky}},\ }\href@noop {} {\bibfield
  {journal} {\bibinfo  {journal} {Nature}\ }\textbf {\bibinfo {volume} {534}},\
  \bibinfo {pages} {122} (\bibinfo {year} {2016})}\BibitemShut {NoStop}%
\bibitem [{\citenamefont {McGuyer}\ \emph {et~al.}(2013)\citenamefont
  {McGuyer}, \citenamefont {Osborn}, \citenamefont {McDonald}, \citenamefont
  {Reinaudi}, \citenamefont {Skomorowski}, \citenamefont {Moszynski},\ and\
  \citenamefont {Zelevinsky}}]{mcguyer2013nonadiabatic}%
  \BibitemOpen
  \bibfield  {author} {\bibinfo {author} {\bibfnamefont {B.}~\bibnamefont
  {McGuyer}}, \bibinfo {author} {\bibfnamefont {C.}~\bibnamefont {Osborn}},
  \bibinfo {author} {\bibfnamefont {M.}~\bibnamefont {McDonald}}, \bibinfo
  {author} {\bibfnamefont {G.}~\bibnamefont {Reinaudi}}, \bibinfo {author}
  {\bibfnamefont {W.}~\bibnamefont {Skomorowski}}, \bibinfo {author}
  {\bibfnamefont {R.}~\bibnamefont {Moszynski}}, \ and\ \bibinfo {author}
  {\bibfnamefont {T.}~\bibnamefont {Zelevinsky}},\ }\href@noop {} {\bibfield
  {journal} {\bibinfo  {journal} {Phys. Rev. Lett.}\ }\textbf {\bibinfo
  {volume} {111}},\ \bibinfo {pages} {243003} (\bibinfo {year}
  {2013})}\BibitemShut {NoStop}%
\bibitem [{\citenamefont {McGuyer}\ \emph
  {et~al.}(2015{\natexlab{b}})\citenamefont {McGuyer}, \citenamefont
  {McDonald}, \citenamefont {Iwata}, \citenamefont {Skomorowski}, \citenamefont
  {Moszynski},\ and\ \citenamefont {Zelevinsky}}]{mcguyer2015control}%
  \BibitemOpen
  \bibfield  {author} {\bibinfo {author} {\bibfnamefont {B.}~\bibnamefont
  {McGuyer}}, \bibinfo {author} {\bibfnamefont {M.}~\bibnamefont {McDonald}},
  \bibinfo {author} {\bibfnamefont {G.}~\bibnamefont {Iwata}}, \bibinfo
  {author} {\bibfnamefont {W.}~\bibnamefont {Skomorowski}}, \bibinfo {author}
  {\bibfnamefont {R.}~\bibnamefont {Moszynski}}, \ and\ \bibinfo {author}
  {\bibfnamefont {T.}~\bibnamefont {Zelevinsky}},\ }\href@noop {} {\bibfield
  {journal} {\bibinfo  {journal} {Phys. Rev. Lett.}\ }\textbf {\bibinfo
  {volume} {115}},\ \bibinfo {pages} {053001} (\bibinfo {year}
  {2015}{\natexlab{b}})}\BibitemShut {NoStop}%
\bibitem [{\citenamefont {Tomza}\ \emph {et~al.}(2012)\citenamefont {Tomza},
  \citenamefont {Goerz}, \citenamefont {Musia{\l}}, \citenamefont {Moszynski},\
  and\ \citenamefont {Koch}}]{tomza2012optimized}%
  \BibitemOpen
  \bibfield  {author} {\bibinfo {author} {\bibfnamefont {M.}~\bibnamefont
  {Tomza}}, \bibinfo {author} {\bibfnamefont {M.~H.}\ \bibnamefont {Goerz}},
  \bibinfo {author} {\bibfnamefont {M.}~\bibnamefont {Musia{\l}}}, \bibinfo
  {author} {\bibfnamefont {R.}~\bibnamefont {Moszynski}}, \ and\ \bibinfo
  {author} {\bibfnamefont {C.~P.}\ \bibnamefont {Koch}},\ }\href@noop {}
  {\bibfield  {journal} {\bibinfo  {journal} {Phys. Rev. A}\ }\textbf {\bibinfo
  {volume} {86}},\ \bibinfo {pages} {043424} (\bibinfo {year}
  {2012})}\BibitemShut {NoStop}%
\bibitem [{\citenamefont {Skomorowski}\ and\ \citenamefont
  {Moszynski}(2011)}]{skomorowski2011long}%
  \BibitemOpen
  \bibfield  {author} {\bibinfo {author} {\bibfnamefont {W.}~\bibnamefont
  {Skomorowski}}\ and\ \bibinfo {author} {\bibfnamefont {R.}~\bibnamefont
  {Moszynski}},\ }\href@noop {} {\bibfield  {journal} {\bibinfo  {journal} {J.
  Chem. Phys.}\ }\textbf {\bibinfo {volume} {134}},\ \bibinfo {pages} {124117}
  (\bibinfo {year} {2011})}\BibitemShut {NoStop}%
\bibitem [{\citenamefont {Tomza}\ \emph {et~al.}(2013)\citenamefont {Tomza},
  \citenamefont {Skomorowski}, \citenamefont {Musia{\l}}, \citenamefont
  {Gonz{\'a}lez-F{\'e}rez}, \citenamefont {Koch},\ and\ \citenamefont
  {Moszynski}}]{tomza2013interatomic}%
  \BibitemOpen
  \bibfield  {author} {\bibinfo {author} {\bibfnamefont {M.}~\bibnamefont
  {Tomza}}, \bibinfo {author} {\bibfnamefont {W.}~\bibnamefont {Skomorowski}},
  \bibinfo {author} {\bibfnamefont {M.}~\bibnamefont {Musia{\l}}}, \bibinfo
  {author} {\bibfnamefont {R.}~\bibnamefont {Gonz{\'a}lez-F{\'e}rez}}, \bibinfo
  {author} {\bibfnamefont {C.~P.}\ \bibnamefont {Koch}}, \ and\ \bibinfo
  {author} {\bibfnamefont {R.}~\bibnamefont {Moszynski}},\ }\href@noop {}
  {\bibfield  {journal} {\bibinfo  {journal} {Mol. Phys.}\ }\textbf {\bibinfo
  {volume} {111}},\ \bibinfo {pages} {1781} (\bibinfo {year}
  {2013})}\BibitemShut {NoStop}%
\bibitem [{\citenamefont {Koch}\ and\ \citenamefont
  {J{\o}rgensen}(1990)}]{koch1990coupled}%
  \BibitemOpen
  \bibfield  {author} {\bibinfo {author} {\bibfnamefont {H.}~\bibnamefont
  {Koch}}\ and\ \bibinfo {author} {\bibfnamefont {P.}~\bibnamefont
  {J{\o}rgensen}},\ }\href {\doibase 10.1063/1.458814} {\bibfield  {journal}
  {\bibinfo  {journal} {J. Chem. Phys.}\ }\textbf {\bibinfo {volume} {93}},\
  \bibinfo {pages} {3333} (\bibinfo {year} {1990})}\BibitemShut {NoStop}%
\bibitem [{\citenamefont {Pedersen}\ and\ \citenamefont
  {Koch}(1997)}]{koch1997coupled}%
  \BibitemOpen
  \bibfield  {author} {\bibinfo {author} {\bibfnamefont {T.~B.}\ \bibnamefont
  {Pedersen}}\ and\ \bibinfo {author} {\bibfnamefont {H.}~\bibnamefont
  {Koch}},\ }\href {\doibase 10.1063/1.473814} {\bibfield  {journal} {\bibinfo
  {journal} {J. Chem. Phys.}\ }\textbf {\bibinfo {volume} {106}},\ \bibinfo
  {pages} {8059} (\bibinfo {year} {1997})}\BibitemShut {NoStop}%
\bibitem [{\citenamefont {Drake}(2006)}]{drake2006springer}%
  \BibitemOpen
  \bibfield  {author} {\bibinfo {author} {\bibfnamefont {G.~W.}\ \bibnamefont
  {Drake}},\ }\href@noop {} {\emph {\bibinfo {title} {{Springer Handbook of
  Atomic, Molecular, and Optical Physics}}}}\ (\bibinfo  {publisher} {Springer,
  New York},\ \bibinfo {year} {2006})\BibitemShut {NoStop}%
\bibitem [{\citenamefont {Koch}\ \emph {et~al.}(1994)\citenamefont {Koch},
  \citenamefont {Kobayashi}, \citenamefont {de~Meras},\ and\ \citenamefont
  {J{\o}rgensen}}]{koch1994caclculation}%
  \BibitemOpen
  \bibfield  {author} {\bibinfo {author} {\bibfnamefont {H.}~\bibnamefont
  {Koch}}, \bibinfo {author} {\bibfnamefont {R.}~\bibnamefont {Kobayashi}},
  \bibinfo {author} {\bibfnamefont {A.~S.}\ \bibnamefont {de~Meras}}, \ and\
  \bibinfo {author} {\bibfnamefont {P.}~\bibnamefont {J{\o}rgensen}},\ }\href
  {\doibase 10.1063/1.466321} {\bibfield  {journal} {\bibinfo  {journal} {J.
  Chem. Phys.}\ }\textbf {\bibinfo {volume} {100}},\ \bibinfo {pages} {4393}
  (\bibinfo {year} {1994})}\BibitemShut {NoStop}%
\bibitem [{\citenamefont {Christiansen}, \citenamefont {J{\o}rgensen},\ and\
  \citenamefont {H{\"a}ttig}(1998)}]{christiansen1998response}%
  \BibitemOpen
  \bibfield  {author} {\bibinfo {author} {\bibfnamefont {O.}~\bibnamefont
  {Christiansen}}, \bibinfo {author} {\bibfnamefont {P.}~\bibnamefont
  {J{\o}rgensen}}, \ and\ \bibinfo {author} {\bibfnamefont {C.}~\bibnamefont
  {H{\"a}ttig}},\ }\href {\doibase
  10.1002/(SICI)1097-461X(1998)68:1<1::AID-QUA1>3.0.CO;2-Z} {\bibfield
  {journal} {\bibinfo  {journal} {Int. J. Quant. Chem.}\ }\textbf {\bibinfo
  {volume} {68}},\ \bibinfo {pages} {1} (\bibinfo {year} {1998})}\BibitemShut
  {NoStop}%
\bibitem [{\citenamefont {Tucholska}, \citenamefont {Modrzejewski},\ and\
  \citenamefont {Moszynski}(2014)}]{tucholska2014transition}%
  \BibitemOpen
  \bibfield  {author} {\bibinfo {author} {\bibfnamefont {A.~M.}\ \bibnamefont
  {Tucholska}}, \bibinfo {author} {\bibfnamefont {M.}~\bibnamefont
  {Modrzejewski}}, \ and\ \bibinfo {author} {\bibfnamefont {R.}~\bibnamefont
  {Moszynski}},\ }\href@noop {} {\bibfield  {journal} {\bibinfo  {journal} {J.
  Chem. Phys.}\ }\textbf {\bibinfo {volume} {141}},\ \bibinfo {pages} {124109}
  (\bibinfo {year} {2014})}\BibitemShut {NoStop}%
\bibitem [{\citenamefont {Moszynski}, \citenamefont {Żuchowski},\ and\
  \citenamefont {Jeziorski}(2005)}]{moszynski2005time}%
  \BibitemOpen
  \bibfield  {author} {\bibinfo {author} {\bibfnamefont {R.}~\bibnamefont
  {Moszynski}}, \bibinfo {author} {\bibfnamefont {P.~S.}\ \bibnamefont
  {Żuchowski}}, \ and\ \bibinfo {author} {\bibfnamefont {B.}~\bibnamefont
  {Jeziorski}},\ }\href {\doibase 10.1135/cccc20051109} {\bibfield  {journal}
  {\bibinfo  {journal} {Coll. Czech. Chem. Commun}\ }\textbf {\bibinfo {volume}
  {70}},\ \bibinfo {pages} {1109} (\bibinfo {year} {2005})}\BibitemShut
  {NoStop}%
\bibitem [{\citenamefont {Paw{\l}owski}, \citenamefont {Olsen},\ and\
  \citenamefont {J{\o}rgensen}(2015)}]{pawlowski2015molecular}%
  \BibitemOpen
  \bibfield  {author} {\bibinfo {author} {\bibfnamefont {F.}~\bibnamefont
  {Paw{\l}owski}}, \bibinfo {author} {\bibfnamefont {J.}~\bibnamefont {Olsen}},
  \ and\ \bibinfo {author} {\bibfnamefont {P.}~\bibnamefont {J{\o}rgensen}},\
  }\href@noop {} {\bibfield  {journal} {\bibinfo  {journal} {J. Chem. Phys.}\
  }\textbf {\bibinfo {volume} {142}},\ \bibinfo {pages} {114109} (\bibinfo
  {year} {2015})}\BibitemShut {NoStop}%
\bibitem [{\citenamefont {Jeziorski}\ and\ \citenamefont
  {Moszynski}(1993)}]{jeziorski1993explicitly}%
  \BibitemOpen
  \bibfield  {author} {\bibinfo {author} {\bibfnamefont {B.}~\bibnamefont
  {Jeziorski}}\ and\ \bibinfo {author} {\bibfnamefont {R.}~\bibnamefont
  {Moszynski}},\ }\href {\doibase 10.1002/qua.560480303} {\bibfield  {journal}
  {\bibinfo  {journal} {Int. J. Quant. Chem.}\ }\textbf {\bibinfo {volume}
  {48}},\ \bibinfo {pages} {161} (\bibinfo {year} {1993})}\BibitemShut
  {NoStop}%
\bibitem [{\citenamefont {Koch}\ \emph {et~al.}(1997)\citenamefont {Koch},
  \citenamefont {Christiansen}, \citenamefont {J{\o}rgensen}, \citenamefont
  {{Sanchez De Merás}},\ and\ \citenamefont {Helgaker}}]{koch1997cc3}%
  \BibitemOpen
  \bibfield  {author} {\bibinfo {author} {\bibfnamefont {H.}~\bibnamefont
  {Koch}}, \bibinfo {author} {\bibfnamefont {O.}~\bibnamefont {Christiansen}},
  \bibinfo {author} {\bibfnamefont {P.}~\bibnamefont {J{\o}rgensen}}, \bibinfo
  {author} {\bibfnamefont {A.~M.}\ \bibnamefont {{Sanchez De Merás}}}, \ and\
  \bibinfo {author} {\bibfnamefont {T.}~\bibnamefont {Helgaker}},\ }\href
  {\doibase 10.1063/1.473322} {\bibfield  {journal} {\bibinfo  {journal} {J.
  Chem. Phys.}\ }\textbf {\bibinfo {volume} {106}},\ \bibinfo {pages} {1808}
  (\bibinfo {year} {1997})}\BibitemShut {NoStop}%
\bibitem [{\citenamefont {Korona}, \citenamefont {Przybytek},\ and\
  \citenamefont {Jeziorski}(2006)}]{korona2006time}%
  \BibitemOpen
  \bibfield  {author} {\bibinfo {author} {\bibfnamefont {T.}~\bibnamefont
  {Korona}}, \bibinfo {author} {\bibfnamefont {M.}~\bibnamefont {Przybytek}}, \
  and\ \bibinfo {author} {\bibfnamefont {B.}~\bibnamefont {Jeziorski}},\ }\href
  {\doibase 10.1080/00268970600673975} {\bibfield  {journal} {\bibinfo
  {journal} {Mol. Phys.}\ }\textbf {\bibinfo {volume} {104}},\ \bibinfo {pages}
  {2303} (\bibinfo {year} {2006})}\BibitemShut {NoStop}%
\bibitem [{\citenamefont {Helgaker}, \citenamefont {Jorgensen},\ and\
  \citenamefont {Olsen}(2013)}]{helgaker2013molecular}%
  \BibitemOpen
  \bibfield  {author} {\bibinfo {author} {\bibfnamefont {T.}~\bibnamefont
  {Helgaker}}, \bibinfo {author} {\bibfnamefont {P.}~\bibnamefont {Jorgensen}},
  \ and\ \bibinfo {author} {\bibfnamefont {J.}~\bibnamefont {Olsen}},\
  }\href@noop {} {\emph {\bibinfo {title} {Molecular electronic-structure
  theory}}}\ (\bibinfo  {publisher} {Wiley, New York},\ \bibinfo {year}
  {2013})\BibitemShut {NoStop}%
\bibitem [{\citenamefont {Dunning~Jr}(1989)}]{dunning1989gaussian}%
  \BibitemOpen
  \bibfield  {author} {\bibinfo {author} {\bibfnamefont {T.~H.}\ \bibnamefont
  {Dunning~Jr}},\ }\href@noop {} {\bibfield  {journal} {\bibinfo  {journal} {J.
  Chem. Phys.}\ }\textbf {\bibinfo {volume} {90}},\ \bibinfo {pages} {1007}
  (\bibinfo {year} {1989})}\BibitemShut {NoStop}%
\bibitem [{\citenamefont {Lesiuk}\ \emph {et~al.}(2015)\citenamefont {Lesiuk},
  \citenamefont {Przybytek}, \citenamefont {Musia\l{}}, \citenamefont
  {Jeziorski},\ and\ \citenamefont {Moszynski}}]{lesiukIII}%
  \BibitemOpen
  \bibfield  {author} {\bibinfo {author} {\bibfnamefont {M.}~\bibnamefont
  {Lesiuk}}, \bibinfo {author} {\bibfnamefont {M.}~\bibnamefont {Przybytek}},
  \bibinfo {author} {\bibfnamefont {M.}~\bibnamefont {Musia\l{}}}, \bibinfo
  {author} {\bibfnamefont {B.}~\bibnamefont {Jeziorski}}, \ and\ \bibinfo
  {author} {\bibfnamefont {R.}~\bibnamefont {Moszynski}},\ }\href {\doibase
  10.1103/PhysRevA.91.012510} {\bibfield  {journal} {\bibinfo  {journal} {Phys.
  Rev. A}\ }\textbf {\bibinfo {volume} {91}},\ \bibinfo {pages} {012510}
  (\bibinfo {year} {2015})}\BibitemShut {NoStop}%
\bibitem [{\citenamefont {Feller}(1996)}]{feller1996role}%
  \BibitemOpen
  \bibfield  {author} {\bibinfo {author} {\bibfnamefont {D.}~\bibnamefont
  {Feller}},\ }\href {\doibase
  10.1002/(SICI)1096-987X(199610)17:13<1571::AID-JCC9>3.0.CO;2-P} {\bibfield
  {journal} {\bibinfo  {journal} {J. Comp. Chem.}\ }\textbf {\bibinfo {volume}
  {17}},\ \bibinfo {pages} {1571} (\bibinfo {year} {1996})}\BibitemShut
  {NoStop}%
\bibitem [{\citenamefont {Schuchardt}\ \emph {et~al.}(2007)\citenamefont
  {Schuchardt}, \citenamefont {Didier}, \citenamefont {Elsethagen},
  \citenamefont {Sun}, \citenamefont {Gurumoorthi}, \citenamefont {Chase},
  \citenamefont {Li},\ and\ \citenamefont {Windus}}]{schuchardt2007basis}%
  \BibitemOpen
  \bibfield  {author} {\bibinfo {author} {\bibfnamefont {K.~L.}\ \bibnamefont
  {Schuchardt}}, \bibinfo {author} {\bibfnamefont {B.~T.}\ \bibnamefont
  {Didier}}, \bibinfo {author} {\bibfnamefont {T.}~\bibnamefont {Elsethagen}},
  \bibinfo {author} {\bibfnamefont {L.}~\bibnamefont {Sun}}, \bibinfo {author}
  {\bibfnamefont {V.}~\bibnamefont {Gurumoorthi}}, \bibinfo {author}
  {\bibfnamefont {J.}~\bibnamefont {Chase}}, \bibinfo {author} {\bibfnamefont
  {J.}~\bibnamefont {Li}}, \ and\ \bibinfo {author} {\bibfnamefont {T.~L.}\
  \bibnamefont {Windus}},\ }\href {\doibase 10.1021/ci600510j} {\bibfield
  {journal} {\bibinfo  {journal} {J. Chem. Inf. Model.}\ }\textbf {\bibinfo
  {volume} {47}},\ \bibinfo {pages} {1045} (\bibinfo {year}
  {2007})}\BibitemShut {NoStop}%
\bibitem [{\citenamefont {Lesiuk}\ and\ \citenamefont
  {Moszynski}(2014{\natexlab{a}})}]{lesiukI}%
  \BibitemOpen
  \bibfield  {author} {\bibinfo {author} {\bibfnamefont {M.}~\bibnamefont
  {Lesiuk}}\ and\ \bibinfo {author} {\bibfnamefont {R.}~\bibnamefont
  {Moszynski}},\ }\href {\doibase 10.1103/PhysRevE.90.063318} {\bibfield
  {journal} {\bibinfo  {journal} {Phys. Rev. E}\ }\textbf {\bibinfo {volume}
  {90}},\ \bibinfo {pages} {063318} (\bibinfo {year}
  {2014}{\natexlab{a}})}\BibitemShut {NoStop}%
\bibitem [{\citenamefont {Lesiuk}\ and\ \citenamefont
  {Moszynski}(2014{\natexlab{b}})}]{lesiukII}%
  \BibitemOpen
  \bibfield  {author} {\bibinfo {author} {\bibfnamefont {M.}~\bibnamefont
  {Lesiuk}}\ and\ \bibinfo {author} {\bibfnamefont {R.}~\bibnamefont
  {Moszynski}},\ }\href {\doibase 10.1103/PhysRevE.90.063319} {\bibfield
  {journal} {\bibinfo  {journal} {Phys. Rev. E}\ }\textbf {\bibinfo {volume}
  {90}},\ \bibinfo {pages} {063319} (\bibinfo {year}
  {2014}{\natexlab{b}})}\BibitemShut {NoStop}%
\bibitem [{\citenamefont {Aidas}\ \emph {et~al.}(2014)\citenamefont {Aidas},
  \citenamefont {Angeli}, \citenamefont {Bak}, \citenamefont {Bakken},
  \citenamefont {Bast}, \citenamefont {Boman}, \citenamefont {Christiansen},
  \citenamefont {Cimiraglia}, \citenamefont {Coriani}, \citenamefont {Dahle}
  \emph {et~al.}}]{aidas2014dalton}%
  \BibitemOpen
  \bibfield  {author} {\bibinfo {author} {\bibfnamefont {K.}~\bibnamefont
  {Aidas}}, \bibinfo {author} {\bibfnamefont {C.}~\bibnamefont {Angeli}},
  \bibinfo {author} {\bibfnamefont {K.~L.}\ \bibnamefont {Bak}}, \bibinfo
  {author} {\bibfnamefont {V.}~\bibnamefont {Bakken}}, \bibinfo {author}
  {\bibfnamefont {R.}~\bibnamefont {Bast}}, \bibinfo {author} {\bibfnamefont
  {L.}~\bibnamefont {Boman}}, \bibinfo {author} {\bibfnamefont
  {O.}~\bibnamefont {Christiansen}}, \bibinfo {author} {\bibfnamefont
  {R.}~\bibnamefont {Cimiraglia}}, \bibinfo {author} {\bibfnamefont
  {S.}~\bibnamefont {Coriani}}, \bibinfo {author} {\bibfnamefont
  {P.}~\bibnamefont {Dahle}},  \emph {et~al.},\ }\href@noop {} {\bibfield
  {journal} {\bibinfo  {journal} {WIREs: Comp. Mol. Sci.}\ }\textbf {\bibinfo
  {volume} {4}},\ \bibinfo {pages} {269} (\bibinfo {year} {2014})}\BibitemShut
  {NoStop}%
\bibitem [{\citenamefont {Fischer}(1975)}]{fischer1975theoretical}%
  \BibitemOpen
  \bibfield  {author} {\bibinfo {author} {\bibfnamefont {C.~F.}\ \bibnamefont
  {Fischer}},\ }\href@noop {} {\bibfield  {journal} {\bibinfo  {journal} {Can.
  J. Phys.}\ }\textbf {\bibinfo {volume} {53}},\ \bibinfo {pages} {338}
  (\bibinfo {year} {1975})}\BibitemShut {NoStop}%
\bibitem [{\citenamefont {Chang}(1986)}]{chang1986effect}%
  \BibitemOpen
  \bibfield  {author} {\bibinfo {author} {\bibfnamefont {T.}~\bibnamefont
  {Chang}},\ }\href@noop {} {\bibfield  {journal} {\bibinfo  {journal} {Phys.
  Rev. A}\ }\textbf {\bibinfo {volume} {34}},\ \bibinfo {pages} {4550}
  (\bibinfo {year} {1986})}\BibitemShut {NoStop}%
\bibitem [{\citenamefont {Zheng}\ \emph {et~al.}(2001)\citenamefont {Zheng},
  \citenamefont {Wang}, \citenamefont {Yang}, \citenamefont {Zhou},
  \citenamefont {Ma}, \citenamefont {Wu},\ and\ \citenamefont
  {Xu}}]{zheng2001transition}%
  \BibitemOpen
  \bibfield  {author} {\bibinfo {author} {\bibfnamefont {N.~W.}\ \bibnamefont
  {Zheng}}, \bibinfo {author} {\bibfnamefont {T.}~\bibnamefont {Wang}},
  \bibinfo {author} {\bibfnamefont {R.~Y.}\ \bibnamefont {Yang}}, \bibinfo
  {author} {\bibfnamefont {T.}~\bibnamefont {Zhou}}, \bibinfo {author}
  {\bibfnamefont {D.~X.}\ \bibnamefont {Ma}}, \bibinfo {author} {\bibfnamefont
  {Y.~G.}\ \bibnamefont {Wu}}, \ and\ \bibinfo {author} {\bibfnamefont {H.~T.}\
  \bibnamefont {Xu}},\ }\href@noop {} {\bibfield  {journal} {\bibinfo
  {journal} {At. Data Nucl. Data Tables}\ }\textbf {\bibinfo {volume} {79}},\
  \bibinfo {pages} {109} (\bibinfo {year} {2001})}\BibitemShut {NoStop}%
\bibitem [{\citenamefont {Gratton}\ \emph {et~al.}(2003)\citenamefont
  {Gratton}, \citenamefont {Carretta}, \citenamefont {Claudi}, \citenamefont
  {Lucatello},\ and\ \citenamefont {Barbieri}}]{gratton2003abundances}%
  \BibitemOpen
  \bibfield  {author} {\bibinfo {author} {\bibfnamefont {R.~G.}\ \bibnamefont
  {Gratton}}, \bibinfo {author} {\bibfnamefont {E.}~\bibnamefont {Carretta}},
  \bibinfo {author} {\bibfnamefont {R.}~\bibnamefont {Claudi}}, \bibinfo
  {author} {\bibfnamefont {S.}~\bibnamefont {Lucatello}}, \ and\ \bibinfo
  {author} {\bibfnamefont {M.}~\bibnamefont {Barbieri}},\ }\href@noop {}
  {\bibfield  {journal} {\bibinfo  {journal} {A \& A}\ }\textbf {\bibinfo
  {volume} {404}},\ \bibinfo {pages} {187} (\bibinfo {year}
  {2003})}\BibitemShut {NoStop}%
\bibitem [{\citenamefont {Schaefer}(1971)}]{schaefer1971measured}%
  \BibitemOpen
  \bibfield  {author} {\bibinfo {author} {\bibfnamefont {A.}~\bibnamefont
  {Schaefer}},\ }\href@noop {} {\bibfield  {journal} {\bibinfo  {journal}
  {Astrophys. J.}\ }\textbf {\bibinfo {volume} {163}},\ \bibinfo {pages} {411}
  (\bibinfo {year} {1971})}\BibitemShut {NoStop}%
\bibitem [{\citenamefont {Chantepie}\ \emph {et~al.}(1989)\citenamefont
  {Chantepie}, \citenamefont {Cheron}, \citenamefont {Cojan}, \citenamefont
  {Landais}, \citenamefont {Laniepce}, \citenamefont {Moudden},\ and\
  \citenamefont {Aymar}}]{chantepie1989time}%
  \BibitemOpen
  \bibfield  {author} {\bibinfo {author} {\bibfnamefont {M.}~\bibnamefont
  {Chantepie}}, \bibinfo {author} {\bibfnamefont {B.}~\bibnamefont {Cheron}},
  \bibinfo {author} {\bibfnamefont {J.}~\bibnamefont {Cojan}}, \bibinfo
  {author} {\bibfnamefont {J.}~\bibnamefont {Landais}}, \bibinfo {author}
  {\bibfnamefont {B.}~\bibnamefont {Laniepce}}, \bibinfo {author}
  {\bibfnamefont {A.}~\bibnamefont {Moudden}}, \ and\ \bibinfo {author}
  {\bibfnamefont {M.}~\bibnamefont {Aymar}},\ }\href@noop {} {\bibfield
  {journal} {\bibinfo  {journal} {J. Phys. B}\ }\textbf {\bibinfo {volume}
  {22}},\ \bibinfo {pages} {2377} (\bibinfo {year} {1989})}\BibitemShut
  {NoStop}%
\bibitem [{\citenamefont {J{\"o}nsson}\ \emph {et~al.}(1984)\citenamefont
  {J{\"o}nsson}, \citenamefont {Levinson}, \citenamefont {Persson},\ and\
  \citenamefont {Wahlstr{\"o}m}}]{jonsson1984natural}%
  \BibitemOpen
  \bibfield  {author} {\bibinfo {author} {\bibfnamefont {G.}~\bibnamefont
  {J{\"o}nsson}}, \bibinfo {author} {\bibfnamefont {C.}~\bibnamefont
  {Levinson}}, \bibinfo {author} {\bibfnamefont {A.}~\bibnamefont {Persson}}, \
  and\ \bibinfo {author} {\bibfnamefont {C.-G.}\ \bibnamefont
  {Wahlstr{\"o}m}},\ }\href@noop {} {\bibfield  {journal} {\bibinfo  {journal}
  {Z. Phys. A}\ }\textbf {\bibinfo {volume} {316}},\ \bibinfo {pages} {255}
  (\bibinfo {year} {1984})}\BibitemShut {NoStop}%
\bibitem [{\citenamefont {Aldenius}\ \emph {et~al.}(2007)\citenamefont
  {Aldenius}, \citenamefont {Tanner}, \citenamefont {Johansson}, \citenamefont
  {Lundberg},\ and\ \citenamefont {Ryan}}]{aldenius2007experimental}%
  \BibitemOpen
  \bibfield  {author} {\bibinfo {author} {\bibfnamefont {M.}~\bibnamefont
  {Aldenius}}, \bibinfo {author} {\bibfnamefont {J.~D.}\ \bibnamefont
  {Tanner}}, \bibinfo {author} {\bibfnamefont {S.}~\bibnamefont {Johansson}},
  \bibinfo {author} {\bibfnamefont {H.}~\bibnamefont {Lundberg}}, \ and\
  \bibinfo {author} {\bibfnamefont {S.~G.}\ \bibnamefont {Ryan}},\ }\href@noop
  {} {\bibfield  {journal} {\bibinfo  {journal} {A \& A}\ }\textbf {\bibinfo
  {volume} {461}},\ \bibinfo {pages} {767} (\bibinfo {year}
  {2007})}\BibitemShut {NoStop}%
\bibitem [{\citenamefont {Kwiatkowski}, \citenamefont {Teppner},\ and\
  \citenamefont {Zimmermann}(1980)}]{kwiatkowski1980lifetime}%
  \BibitemOpen
  \bibfield  {author} {\bibinfo {author} {\bibfnamefont {M.}~\bibnamefont
  {Kwiatkowski}}, \bibinfo {author} {\bibfnamefont {U.}~\bibnamefont
  {Teppner}}, \ and\ \bibinfo {author} {\bibfnamefont {P.}~\bibnamefont
  {Zimmermann}},\ }\href@noop {} {\bibfield  {journal} {\bibinfo  {journal} {Z.
  Phys. A}\ }\textbf {\bibinfo {volume} {294}},\ \bibinfo {pages} {109}
  (\bibinfo {year} {1980})}\BibitemShut {NoStop}%
\bibitem [{\citenamefont {Andersen}, \citenamefont {Molhave},\ and\
  \citenamefont {Sorensen}(1972)}]{andersen1972lifetimes}%
  \BibitemOpen
  \bibfield  {author} {\bibinfo {author} {\bibfnamefont {T.}~\bibnamefont
  {Andersen}}, \bibinfo {author} {\bibfnamefont {L.}~\bibnamefont {Molhave}}, \
  and\ \bibinfo {author} {\bibfnamefont {G.}~\bibnamefont {Sorensen}},\
  }\href@noop {} {\bibfield  {journal} {\bibinfo  {journal} {Astrophys. J.}\
  }\textbf {\bibinfo {volume} {178}},\ \bibinfo {pages} {577} (\bibinfo {year}
  {1972})}\BibitemShut {NoStop}%
\bibitem [{\citenamefont {Ueda}, \citenamefont {Karasawa},\ and\ \citenamefont
  {Fukuda}(1982)}]{ueda1982measurements}%
  \BibitemOpen
  \bibfield  {author} {\bibinfo {author} {\bibfnamefont {K.}~\bibnamefont
  {Ueda}}, \bibinfo {author} {\bibfnamefont {M.}~\bibnamefont {Karasawa}}, \
  and\ \bibinfo {author} {\bibfnamefont {K.}~\bibnamefont {Fukuda}},\
  }\href@noop {} {\bibfield  {journal} {\bibinfo  {journal} {J. Phys. Soc.
  Jpn}\ }\textbf {\bibinfo {volume} {51}},\ \bibinfo {pages} {2267} (\bibinfo
  {year} {1982})}\BibitemShut {NoStop}%
\bibitem [{\citenamefont {Havey}, \citenamefont {Balling},\ and\ \citenamefont
  {Wright}(1977)}]{havey1977direct}%
  \BibitemOpen
  \bibfield  {author} {\bibinfo {author} {\bibfnamefont {M.}~\bibnamefont
  {Havey}}, \bibinfo {author} {\bibfnamefont {L.}~\bibnamefont {Balling}}, \
  and\ \bibinfo {author} {\bibfnamefont {J.}~\bibnamefont {Wright}},\
  }\href@noop {} {\bibfield  {journal} {\bibinfo  {journal} {J. Opt. Soc. Am.}\
  }\textbf {\bibinfo {volume} {67}},\ \bibinfo {pages} {488} (\bibinfo {year}
  {1977})}\BibitemShut {NoStop}%
\bibitem [{\citenamefont {Moccia}\ and\ \citenamefont
  {Spizzo}(1988)}]{moccia1988atomic}%
  \BibitemOpen
  \bibfield  {author} {\bibinfo {author} {\bibfnamefont {R.}~\bibnamefont
  {Moccia}}\ and\ \bibinfo {author} {\bibfnamefont {P.}~\bibnamefont
  {Spizzo}},\ }\href@noop {} {\bibfield  {journal} {\bibinfo  {journal} {J.
  Phys. B}\ }\textbf {\bibinfo {volume} {21}},\ \bibinfo {pages} {1133}
  (\bibinfo {year} {1988})}\BibitemShut {NoStop}%
\bibitem [{\citenamefont {Victor}, \citenamefont {Stewart},\ and\ \citenamefont
  {Laughlin}(1976)}]{victor1976oscillator}%
  \BibitemOpen
  \bibfield  {author} {\bibinfo {author} {\bibfnamefont {G.}~\bibnamefont
  {Victor}}, \bibinfo {author} {\bibfnamefont {R.}~\bibnamefont {Stewart}}, \
  and\ \bibinfo {author} {\bibfnamefont {C.}~\bibnamefont {Laughlin}},\
  }\href@noop {} {\bibfield  {journal} {\bibinfo  {journal} {Astrophys. J.
  Suppl. Ser.}\ }\textbf {\bibinfo {volume} {31}},\ \bibinfo {pages} {237}
  (\bibinfo {year} {1976})}\BibitemShut {NoStop}%
\bibitem [{\citenamefont {Mendoza}(1981)}]{mendoza1981term}%
  \BibitemOpen
  \bibfield  {author} {\bibinfo {author} {\bibfnamefont {C.}~\bibnamefont
  {Mendoza}},\ }\href@noop {} {\bibfield  {journal} {\bibinfo  {journal} {J.
  Phys. B}\ }\textbf {\bibinfo {volume} {14}},\ \bibinfo {pages} {397}
  (\bibinfo {year} {1981})}\BibitemShut {NoStop}%
\bibitem [{\citenamefont {Werij}\ \emph {et~al.}(1992)\citenamefont {Werij},
  \citenamefont {Greene}, \citenamefont {Theodosiou},\ and\ \citenamefont
  {Gallagher}}]{werij1992oscillator}%
  \BibitemOpen
  \bibfield  {author} {\bibinfo {author} {\bibfnamefont {H.}~\bibnamefont
  {Werij}}, \bibinfo {author} {\bibfnamefont {C.~H.}\ \bibnamefont {Greene}},
  \bibinfo {author} {\bibfnamefont {C.}~\bibnamefont {Theodosiou}}, \ and\
  \bibinfo {author} {\bibfnamefont {A.}~\bibnamefont {Gallagher}},\ }\href@noop
  {} {\bibfield  {journal} {\bibinfo  {journal} {Phys. Rev. A}\ }\textbf
  {\bibinfo {volume} {46}},\ \bibinfo {pages} {1248} (\bibinfo {year}
  {1992})}\BibitemShut {NoStop}%
\bibitem [{\citenamefont {Hunter}, \citenamefont {Walker},\ and\ \citenamefont
  {Weiss}(1986)}]{hunter1986observation}%
  \BibitemOpen
  \bibfield  {author} {\bibinfo {author} {\bibfnamefont {L.}~\bibnamefont
  {Hunter}}, \bibinfo {author} {\bibfnamefont {W.}~\bibnamefont {Walker}}, \
  and\ \bibinfo {author} {\bibfnamefont {D.}~\bibnamefont {Weiss}},\
  }\href@noop {} {\bibfield  {journal} {\bibinfo  {journal} {Phys. Rev. Lett.}\
  }\textbf {\bibinfo {volume} {56}},\ \bibinfo {pages} {823} (\bibinfo {year}
  {1986})}\BibitemShut {NoStop}%
\bibitem [{\citenamefont {Porsev}\ \emph {et~al.}(2008)\citenamefont {Porsev},
  \citenamefont {Ludlow}, \citenamefont {Boyd},\ and\ \citenamefont
  {Ye}}]{porsev2008determination}%
  \BibitemOpen
  \bibfield  {author} {\bibinfo {author} {\bibfnamefont {S.}~\bibnamefont
  {Porsev}}, \bibinfo {author} {\bibfnamefont {A.~D.}\ \bibnamefont {Ludlow}},
  \bibinfo {author} {\bibfnamefont {M.~M.}\ \bibnamefont {Boyd}}, \ and\
  \bibinfo {author} {\bibfnamefont {J.}~\bibnamefont {Ye}},\ }\href@noop {}
  {\bibfield  {journal} {\bibinfo  {journal} {Phys. Rev. A}\ }\textbf {\bibinfo
  {volume} {78}},\ \bibinfo {pages} {032508} (\bibinfo {year}
  {2008})}\BibitemShut {NoStop}%
\bibitem [{\citenamefont {Brinkmann}\ \emph {et~al.}(1969)\citenamefont
  {Brinkmann}, \citenamefont {Goschler}, \citenamefont {Steudel},\ and\
  \citenamefont {Walther}}]{brinkmann1969experimente}%
  \BibitemOpen
  \bibfield  {author} {\bibinfo {author} {\bibfnamefont {U.}~\bibnamefont
  {Brinkmann}}, \bibinfo {author} {\bibfnamefont {J.}~\bibnamefont {Goschler}},
  \bibinfo {author} {\bibfnamefont {A.}~\bibnamefont {Steudel}}, \ and\
  \bibinfo {author} {\bibfnamefont {H.}~\bibnamefont {Walther}},\ }\href@noop
  {} {\bibfield  {journal} {\bibinfo  {journal} {Z. Phys.}\ }\textbf {\bibinfo
  {volume} {228}},\ \bibinfo {pages} {427} (\bibinfo {year}
  {1969})}\BibitemShut {NoStop}%
\bibitem [{\citenamefont {Borisov}, \citenamefont {P.},\ and\ \citenamefont
  {Redko}(1987)}]{borisov87}%
  \BibitemOpen
  \bibfield  {author} {\bibinfo {author} {\bibfnamefont {E.~N.}\ \bibnamefont
  {Borisov}}, \bibinfo {author} {\bibfnamefont {P.~N.}\ \bibnamefont {P.}}, \
  and\ \bibinfo {author} {\bibfnamefont {T.~P.}\ \bibnamefont {Redko}},\
  }\href@noop {} {\bibfield  {journal} {\bibinfo  {journal} {Opt. Spectrosc.}\
  }\textbf {\bibinfo {volume} {63}},\ \bibinfo {pages} {475} (\bibinfo {year}
  {1987})}\BibitemShut {NoStop}%
\bibitem [{\citenamefont {Miller}\ \emph {et~al.}(1992)\citenamefont {Miller},
  \citenamefont {You}, \citenamefont {Cooper},\ and\ \citenamefont
  {Gallagher}}]{miller1992collisional}%
  \BibitemOpen
  \bibfield  {author} {\bibinfo {author} {\bibfnamefont {D.}~\bibnamefont
  {Miller}}, \bibinfo {author} {\bibfnamefont {L.}~\bibnamefont {You}},
  \bibinfo {author} {\bibfnamefont {J.}~\bibnamefont {Cooper}}, \ and\ \bibinfo
  {author} {\bibfnamefont {A.}~\bibnamefont {Gallagher}},\ }\href@noop {}
  {\bibfield  {journal} {\bibinfo  {journal} {Phys. Rev. A}\ }\textbf {\bibinfo
  {volume} {46}},\ \bibinfo {pages} {1303} (\bibinfo {year}
  {1992})}\BibitemShut {NoStop}%
\bibitem [{\citenamefont {Vaeck}, \citenamefont {Godefroid}\ \emph
  {et~al.}(1988)\citenamefont {Vaeck}, \citenamefont {Godefroid} \emph
  {et~al.}}]{vaeck1988multiconfiguration}%
  \BibitemOpen
  \bibfield  {author} {\bibinfo {author} {\bibfnamefont {N.}~\bibnamefont
  {Vaeck}}, \bibinfo {author} {\bibfnamefont {M.}~\bibnamefont {Godefroid}},
  \emph {et~al.},\ }\href@noop {} {\bibfield  {journal} {\bibinfo  {journal}
  {Phys. Rev. A}\ }\textbf {\bibinfo {volume} {38}},\ \bibinfo {pages} {2830}
  (\bibinfo {year} {1988})}\BibitemShut {NoStop}%
\bibitem [{\citenamefont {Lim}, \citenamefont {Stoll},\ and\ \citenamefont
  {Schwerdtfeger}(2006)}]{lim2006relativistic}%
  \BibitemOpen
  \bibfield  {author} {\bibinfo {author} {\bibfnamefont {I.~S.}\ \bibnamefont
  {Lim}}, \bibinfo {author} {\bibfnamefont {H.}~\bibnamefont {Stoll}}, \ and\
  \bibinfo {author} {\bibfnamefont {P.}~\bibnamefont {Schwerdtfeger}},\ }\href
  {\doibase 10.1063/1.2148945} {\bibfield  {journal} {\bibinfo  {journal} {J.
  Chem. Phys.}\ }\textbf {\bibinfo {volume} {124}},\ \bibinfo {pages} {034107}
  (\bibinfo {year} {2006})}\BibitemShut {NoStop}%
\end{thebibliography}%

\end{document}